# High-fidelity simulations of shock initiation of an energetic crystal-binder system due to flyer impact

*Shobhan Roy[1], Pradeep K. Seshadri[1], Chukwudubem Okafor[1], Belinda P. Johnson[2], H. S. Udaykumar[1]\**

[1]*Department of Mechanical Engineering, The University of Iowa, Iowa City, IA 52242*

[2]*Shock and Detonation Physics (M-9), Los Alamos National Laboratory, NM 87545*

*\*corresponding author email: hs-kumar@uiowa.edu*

**Abstract**

Meso-scale simulations of energy localization at hotspots provide closure models for multiscale frameworks of shock-to-detonation transition (SDT). Validation of such meso-scale calculations is challenging as direct comparison with experiments is constrained both by limitations of data acquisition in the experiments (e.g., of temperature fields) and modeling over-simplifications in the simulations. To address the latter problem and bring modeling closer to experiments, we advance a high-fidelity meso-scale computational framework for interface-resolved reactive calculations of shock initiation in plastic-bonded explosives (PBXs). Accurate resolution of shock and interfacial dynamics is achieved through higher-order (5$^{th}$-order WENO) schemes, and sharp interface treatments are implemented for physically accurate material-material interactions. Recently obtained atomistics-consistent material models are used for HMX, with the grid resolution taken down to atomistic scale ($O$(nm)). The crystal geometries are obtained directly from experiments via nano-CT imaging. The impacting flyer plate, energetic crystal and binder are tracked as distinct phases, and flyer-binder impact and separation are simulated, capturing the flyer deformation and the effects of relief waves from the flyer surface. By combining these high-fidelity modeling components, we evaluate how closely simulations can approach experimental data. Overall, this work provides an assessment of which aspects of numerical treatment and material modeling have the greatest impact on meso-scale simulations of flyer-induced initiation of PBXs, and points to where further improvements are necessary.

-------------------------------------------------------------------------------------------------------------------

## 1. Introduction

Shock-to-detonation transition (SDT) in composite energetic materials such as plastic-bonded explosives (PBX) occur over a wide range of length and time scales [2] necessitating multi-scale modeling techniques for predictive calculations. Heterogenous energetic materials (HEs) display strong energy localization at material interfaces [3, 4]. Such localizations – called 'hotspots' – occur at small length/time scales that are difficult to observe experimentally [5-7]. The small length and time scales also make the numerical modeling of shock initiation challenging [8]. Several modeling techniques have been used to simulate the meso-scale (grain-scale) phenomena associated with hotspot formation and growth, with inherent assumptions or simplifications to ensure computational tractability[9-18]. However, modeling over-simplifications at the meso-scale level can percolate errors and uncertainties[19] up to larger length/time scales in a multi-scale framework and yield inaccurate system-level predictions. To develop reliable sub-grid models for macroscale predictive calculations, simulations at the meso-scale must accurately represent: 1) the physics of flyer interaction with the EM sample, 2) the mechanics of crystal-crystal and crystal-binder interactions; and 3) shock interactions with inter- and intra-crystal defects. For physically accurate simulations, the numerical schemes should capture sharp gradients such as shocks, reaction fronts, contact and material discontinuities in the individual bulk materials[20-22] and interface treatments need to account for interaction between media with disparate impedances[8, 23, 24]. In addition, material models must capture the elastoplastic response over a wide range of strain and strain rates[25-28] and reaction chemistry coupled with shocks[29, 30]. The focus of this work is to demonstrate a high-fidelity computational framework that meets these desired conditions, enabling meso-scale simulations to closely adhere to physical flyer-impact-induced shock initiation experiments on model PBX samples[1, 6].

In the context of meso-scale dynamics under shock conditions, a long-standing challenge is to obtain suitable benchmarks to assess the accuracy of simulations. Meso-scale simulations, which pertain to representative volume elements in the µm-mm range and time scales in the ns-µs range, fall into the gap between the ranges of time and spatial scales normally covered by molecular dynamics (MD) on the one hand and experimental observations on the other. MD can inform continuum models at small scales, reaching up to $O(\mu m)$ and $O(ns)$[31, 32], whereas experimental observations are commonly made at the macro-scale, i.e., at or above $O(mm)$ and $O(\mu s)$[33, 34]. One experimental configuration that presents the opportunity to benchmark meso-scale simulations of shock initiation of PBX is the laser-launched flyer impact experiments on micro-geometries conducted by Dlott and coworkers[6, 35]. The tabletop apparatus used for these experiments has been described in previous works[36-40]. Their shock compression microscopy apparatus has demonstrated the ability to capture shock-induced hotspot ignition and growth in model PBX samples with nanosecond time resolution and micrometer spatial resolution, i.e., in the time and space scale ranges pertaining to the meso-scale. High-speed videography enabled the visualization of incipient hotspots, and the temperature of the hotspots was estimated using optical pyrometry. However, since pyrometry provides spatially-averaged measurements, the temperature distributions in the hotspots and their time evolution are not available from these experiments. Nevertheless, the experiments of Dlott et al. [6, 35] do offer quantitative information on transient overall temperature excursions in the deconstructed PBX samples as a result of shock interaction with interfaces and defects. Therefore, these experiments are a good starting point for assessment of meso-scale simulations of flyer-crystal-binder systems relevant to the initiation of PBXs.

In previous meso-scale simulations of the Dlott setup, Roy et al. [1] attempted a head-to-head comparative study between simulations and experiments on flyer-impact induced shock initiation of single-HMX-crystal embedded in binder. The computations were conducted on HMX grains delineated from nano-CT scans of the samples that were tested in experiments. Hotspot temperatures in the crystal at specific time instances in the simulated data were compared with post-processed thermal emission fields obtained from the experiments. Despite several differences between the experimental and simulations setups, Roy et al. [1] reported fair agreement of the peak hotspot temperatures and time-dependent evolution of temperature, though there was clear room for better agreement between the simulations and experiments. It is noted that tandem comparative studies between computations and experiments at the micron length scale like that of Roy et al. [1] are very limited in literature[41, 42], due to both computational and experimental limitations. While the experimental community is striving towards better data acquisition techniques (with respect to measuring the crucial temperature fields around hotspots [43]), improvements in computational models and methodology are also warranted for more reliable comparisons to experiments. In this paper, we present key advances in the simulation framework relative to Roy et al. [1], with particular regard to the three aspects described below.

1.  <u>Updated Atomistics-consistent material models for HMX</u>:
The elastoplastic response of energetic crystals to shock loading is critical in determining the location and intensity of energy localization, and consequently the ignition and growth of hotspots [3, 44, 45]. Isotropic continuum strength models for Eulerian frameworks that predict the plastic deformation of EM are generally of the J2-invariant elastoplastic phenomenological variety, typically calibrated to empirical data [15, 46]. Of the models available in the literature for the yield stress for HMX, some early works used the elastic-perfectly-plastic treatment for the strength of the material[47], assuming a constant yield stress [48-50]. Springer et al. [15] reported an empirically-founded[51-53] Johnson-Cook (JC) model[54, 55] which accounts for strain hardening, strain rate hardening and thermal softening. Other versions of the JC model include those calibrated with MD calculations[56] and single-crystal plasticity calculations[57]. In a recent paper of Nguyen et al. [58], it was shown that calculations using these previous JC models failed to adequately match the hotspot temperature fields and shear band patterns in the pore collapse simulations when compared to MD. Nguyen et al. [58] proposed a modified Johnson-Cook (M-JC) model which yields strain localization and shear band formation; the strain-localization effect was calibrated using all-atom MD calculations[59]. Together with a MD-informed melt curve[60] and pressure-dependent shear modulus[61], an atomistics-consistent, computationally tractable isotropic strength model for the deformation of β-HMX is now available in literature and has been utilized in the current paper.

2. Higher-order numerical accuracy for bulk material and material-material interface treatment:
Most of the meso-scale simulations reported in the literature [62-68] have used low-order accurate (typically nominal $2^{nd}$ order accuracy) numerical schemes. Roy et al. [1] used a $3^{rd}$-order ENO scheme to model shock-interface interaction in energetic material. In this work, a $5^{th}$-order accurate WENO scheme has been employed that has been demonstrated to capture higher level of details in meso- and macro-scale shocked elastoplastic flows[69, 70] for approximately the same computational cost as the $3^{rd}$-order ENO. To capture interface dynamics accurately, the sharp-interface Eulerian levelset tracking used in this work operates in conjunction with a ghost-fluid method (GFM) to impose interfacial conditions between materials. Sharp interfaces and GFM were also employed in Roy et al. [1]. For a more accurate treatment of material-material interfaces, in this work the GFM treatment is advanced[69, 70] to account for the wave interactions at the interface by employing the HLLC approximate Riemann solver[71] to accurately capture the five-wave-structure resulting from elastoplastic wave interactions with a material interface[72]. The calculations here also employ fine mesh resolutions to yield grid independent solutions[73], and to capture material interfaces and flow features in the high-speed flow calculations. In the last section of this paper (Sec. 3.4), a standalone calculation with such high resolution, resolving the fields down to nm scale, has been presented to display the full capability of the numerics and models.

3. Computational setup conforming to experimental impact conditions:
Boundary conditions imposed in the simulations can play an important role in the dynamics of shocks and initiation of hotspots. Modeling impact-induced shock propagation using a velocity or pressure based Dirichlet boundary condition is common in computational works – both a constant velocity/pressure condition to emulate a sustained shock[16, 66, 74-82], and a variable-velocity (pulse) profile for finite-thickness projectile impacts[1, 41, 65, 67, 68, 73]. Some other models intending to approximate an impact condition include initializing the computational domain with a shock near the inlet[11, 83-85] and reverse ballistic impact[57, 86-89]. Only a handful of works have attempted simulating flyer movement, but the effects of the relief wave from the free surface of the flyer are not yet properly captured or reported[15, 42, 90, 91]. In Roy et al. [1], modeling the impact of a 25μm-thick Al flyer through a shock-pulse boundary condition inherently assumed: a) the effective particle velocity ($U_P$) imparted by the impact to the binder surface can be determined, given the impedance mismatch between the metal and polymeric binder, b) the pulse shape and duration emulated the plastic shock followed by the ensuing relief wave (shock wave reflected back from the face of the flyer opposite to the impacting face), and c) the shock is planar in the domain in consideration, i.e., ignoring corner (2D) effects from small-radius flyers and spallation/fragmentation of the flyer and/or binder on impact. In the present work, instead of approximating the flyer dynamics through such idealized boundary treatments, the actual Al flyer impact is simulated in conformity with experiments[6, 35]. Therefore, the wave dynamics, and any departure from the planar assumption due to dimensions of the flyer/binder, are inherently captured by the simulation, more closely representing the experimental setup than in the previous study of Roy et al. [1].

The above three treatments update the meso-scale simulations to the state-of-the-art in terms of numerical accuracy, material model fidelity and resolution on multiprocessor computing systems. We deploy these advances to assess the ability of the simulation framework to capture quantitatively the dynamics of flyer impact induced initiation in a PBX system. To this end, the remainder of the paper is organized as follows. The methodology involving the numerical framework and computational setup is presented briefly in Section 2. In Section 3, simulation results are presented for various cases, systematically assessing the impact of the three different aspects of the advances in the framework listed above. In particular, Section 3.4 presents a high-resolution simulation of a flyer-crystal-binder calculation, demonstrating the high fidelity with which shocks, interfaces and localization zones are captured by the present calculations. Lastly, conclusions are drawn in Section 4.

## 2. Methods

### 2.1 Flow solver

The computational flow solver employed in this work, SCIMITAR3D [65], operates on a fixed Cartesian-grid numerical framework. The governing equations for the solid/fluid flow – conservation relations for mass, momentum, and energy along with the evolution of deviatoric stresses and species mass-fraction – are cast in Eulerian form[92]. These equations are closed using suitable constitutive relations, discretized in space using high-order shock capturing schemes, and explicitly advanced in time for time-accurate solutions to the hyperbolic system[93]. The solver has been benchmarked against numerical[28], experimental[41, 94], and molecular dynamics studies[56, 95], and has been extensively used for simulations of high-strain-rate multi-material reactive flows[50, 73]. Some key aspects of the methodology that are pertinent to the present work have been highlighted in the following subsections. Other details of the overall numerical framework can be found in Supplementary Information and in previous publications[28, 94, 96, 97].

*2.1.1 Numerical schemes for multi-material flow*

The governing equations are discretized using a finite-volume methodology wherein the fluxes are reconstructed using a higher order polynomial - a 5$^{th}$ order WENO-Z scheme[98] implementation that offers low dissipation and prevents oscillations in regions with discontinuities and is suitable for capturing fine details in high-speed flows with shocks[69, 70]. The material boundaries are implicitly tracked as sharp interfaces using the narrow-band levelset method[99] by advecting the signed-distance field along with the flow using a 5$^{th}$ order WENO scheme. The flux reconstruction at multi-material junctions is handled via a modified ghost-fluid method (GFM)[69, 70] wherein the Harten, Lax and van Leer Contact (HLLC)[71] approximate Riemann solver is used to populate the ghost states accounting for each of elastoplastic-elastoplastic, elastoplastic-void, elastoplastic-viscoelastic, viscoelastic-void interactions. The HLLC solver is an improvement over the HLL model[100] and is reported to accurately capture the non-linear wave interactions at material interfaces in high-speed flows [101-103]. It is of special interest in this study to capture material-material interactions accurately, owing to the markedly different shock impedances of the binder and HMX/Aluminum. Traditional contact conditions have been shown to perform poorly in handling the added-mass effect during interactions of materials with large discrepancy in densities and impedances, as seen in fluid-structure interaction applications[23, 104]. Solving a local Riemann problem to populate the ghost cells at the interface automatically takes care of the jump conditions for accurate shock transmission and particle velocity close to the interface; there is no need for explicitly enforcing contact relations customized for specific material interactions. The equation for the evolution of deviatoric stresses is solved using a 2-step operator-splitting method[94] – a first step where pure elastic deformation is assumed and a second step wherein a radial return algorithm[105, 106] is used to remap the predicted stress onto the yield surface. Finally, the solution is advanced in time using a 3$^{rd}$ order total variation diminishing (TVD) Runge-Kutta method[107] while employing an adaptive global time-stepping amenable to temporally resolving the fastest waves in the system. Cumulatively, the numerical framework for multi-material continuum mechanics calculations described above enables accurate capturing of deformed interfaces and wave interactions that occur in the flow regimes investigated in the present work.

*2.1.2. Reaction model*

The chemical decomposition of HMX is modeled using Arrhenius kinetics and consists of a 3-step reaction model due to Tarver[7] that has been widely used in previous works [108-111]. The heat of the reaction is incorporated in the energy conservation relation as a source term. An operator splitting algorithm is used to implicitly integrate the stiff reaction source terms[112]. Unlike reactive burn models for macroscale calculations [89, 113, 114] that employ detonation mixture Hugoniots to get the energy from chemical decomposition, here the shock Hugoniots resulting from the same equation of state are used for the reactants as well as products [11, 115-117].

*2.1.3. Constitutive relations*

1. HMX

Solid β-HMX is modeled in this work as an isotropic elastoplastically deforming material. In previous works[1, 74], the following Johnson-Cook form of yield stress $\sigma_y$ due to Springer et al. (2018)[15] was used:

$$\sigma_y = \left[A + B\varepsilon_{pl}^n\right]\left[1 + C \ln(1 + \dot{\varepsilon}^*)\right]\left[1 - \left(\frac{T - T_{ref}}{T_m - T_{ref}}\right)^M\right]. \tag{1}$$

Here, $\varepsilon_{pl}$ is the equivalent plastic strain, $\dot{\varepsilon}$ is the plastic strain rate, $\dot{\varepsilon}_0$ is a reference plastic strain rate, and $\dot{\varepsilon}^* = \dot{\varepsilon}_{pl}/\dot{\varepsilon}_0$; T is the temperature and $T_{ref}$ is the reference temperature of the material (300 K). A, B, C, n, and M are Johnson-Cook model coefficients and exponents, and can be found in [56]. It was recently shown by Nguyen et al. [58] that the above model displays an unphysical hardening at low shock strength regimes where the majority of the energy localization in hotspots takes place due to shear localization. They proposed a modification to the model, informed by atomistic calculations [118], using which they showed that shear localization at the surface of voids leads to material softening, leading to hotspot shapes and collapse patterns in close agreement with atomistic calculations[58, 87, 119]. This modified Johnson Cook model (M-JC) is as follows: the yield stress is given by the JC formulation in equation (1) below a certain localization criterion. Above the localization criterion defined as: (a) the von-Misses stress is larger than a threshold value, $S_{vM} \geq \sigma_e^c$ (where $S_{vM} = \left\{\frac{3}{2}\left(S_{ij}S_{ij}\right)\right\}^{\frac{1}{2}}$), and (b) the equivalent plastic strain increases above a limit, $\varepsilon_{pl} > \varepsilon_{eq}^c$, the flow-stress model is replaced with:

$$\sigma_y^* = \frac{\sigma_y + \sigma_{flow}^\infty}{2} + \frac{\sigma_{flow}^\infty - \sigma_y}{2}\tanh(\varepsilon_{pl}). \tag{2}$$

In Eq. (2), $\sigma_y$ is computed from Eq. (1). The expressions for $\sigma_e^c$, $\varepsilon_{eq}^c$, and $\sigma_{flow}^\infty$ can be found in Nguyen et al. [58] and are also given in Supplementary Information. The second term on the right-hand side of the M-JC model in Eq. (2) amounts to strain-softening of the shocked crystal in the vicinity of the pore due to localization in shear bands [58]. This softening is driven by the effective plastic strain ($\varepsilon_{pl}$). The form of $\sigma_{flow}^\infty$ used in this relation (see Supplementary Information) is obtained from MD calculations [118]; it accounts for the power-law dependence of the resolved shear stress on the resolved shear-strain rate and for the linear dependence of the resolved flow stress on temperature.

A pressure-dependent model for the shear modulus is used with coefficients calibrated using MD calculations[61]. When the melting temperature is exceeded, HMX is treated as an inviscid liquid (by setting deviatoric stress $(S_{ij}) = 0$); the MD-derived melt curve due to Kroonblawd and Austin [60] is used to denote the transition point for phase change to liquid. The melt-curve, pressure-dependent shear modulus and the M-JC flow stress relation together constitute the atomistics-consistent strength model for β-HMX, hereafter collectively called the M-JC model. This model has been calibrated with atomistic calculations and shown to capture hotspot temperatures due to shear localizations as well as hydrodynamic material jetting when compared head-to-head with MD calculations for a wide range of shock pressures[58].

2. Estane

The binder material used in the computations is the polymer Estane cured with plasticizer (2.5 % Estane®, 2.5 % BDNPA/F), henceforth referred to simply as Estane. The Estane binder used in PBX 9501 has been characterized in [120, 121]. For the present strong shock conditions, Estane is modeled as an inert viscoelastic material for which the stresses are governed by the equation for the evolution of deviatoric stresses given a constant shear modulus of 1 GPa. Beyond a constant melt temperature of 473 K[120], the material is modeled as an inviscid liquid.

The dilatational response of Estane is given by the relation between the pressure P, specific internal energy e, and specific volume ($V = 1/\rho$). In the compressive regime, the relation follows a first-order Mie-Gruneisen equation of state[122] with a reference state of the cold pressure and the specific internal energy at 0:

$$p(e, V) = \Gamma \frac{e}{V} + \begin{cases} \frac{K_0 \vartheta}{(1 - s\vartheta)^2}\left[1 - \frac{\Gamma}{2V}(V - V_0)\right], & \text{if } V \leq V_0 \\ K_0\left(\frac{V_0}{V} - 1\right), & \text{otherwise} \end{cases} \quad (3)$$

In previous works [1, 27, 74], this form was used with an extension to the tensile regime comprising of the pressure being modeled as a linear function of density. However, a further modification to the tensile regime has been adopted in the present work since the robustness of the EOS during expansion is important due to the relief wave transmitted into the binder by the flyer. It was found that regions of the binder can expand beyond its ambient state and modeling the response using an MG EOS calibrated for compression yielded unphysical values of pressure under certain instances of multiple wave reflections in the presence of the crystal. In the modification adopted for the tensile part of the EOS due to Robinson [123], the reference curve is taken to be an isentrope with a single fixed $K_0$, instead of a linear function. Beyond a prescribed tensile cutoff, the isentropic reference curve is further modified as one that does not sustain additional tensile pressure. This modified treatment is observed to be more stable in the presence of pressure fluctuations for the viscoelastic material. The final form is as follows:

$$p(e, V) = \Gamma \frac{e}{V} + \begin{cases} \frac{K_0 \vartheta}{(1 - s\vartheta)^2}\left[1 - \frac{\Gamma V_0}{2V}(\vartheta)\right], & \text{if } \vartheta \geq 0; \\ K_0 \vartheta\left[1 - \frac{\Gamma V_0}{2V}(\vartheta)\right], & \text{if } \vartheta_{min} < \vartheta < 0; \\ K_0 \vartheta_{min}\left[1 - \frac{\Gamma V_0}{2V}(2\vartheta - \vartheta_{min})\right], & \text{if } \vartheta \leq \vartheta_{min}. \end{cases} \quad (4)$$

Here $\vartheta$, K and $\Gamma$ are the volumetric strain, isothermal bulk-modulus and Grüneisen parameter respectively. The subscript "0" refers to the ambient (unstressed) state of the material. $\vartheta_{min}$ is determined by the density at the tensile cutoff state, which for the present calculations, was taken to be $0.85\rho_0$.

3. Aluminum

Aluminum is modeled as an inert elastoplastic material, with a constant melt temperature, shear modulus, specific heat coefficient and thermal conductivity. The Johnson-Cook form given in Eq. (1) is used for yield stress. The Mie-Gruneisen form given in Eq. (3) is used for the EOS. The material-specific coefficients are listed in Supplementary Information.

## 2.2 Computational setup

In the following calculations, three materials are tracked without intermixing, namely Aluminum flyer, Estane binder matrix, and a crystal of β-HMX. The levelset field for each of the materials is tracked separately, with the condition that no two levelsets can have a negative value at the same location, i.e., two materials cannot occupy the same physical space. Collison-separation conditions are applied as described in Rai et al. [124]. The primary computational setup used in this work emulates the thin Al-flyer impact-induced shock initiation experiments conducted by Johnson et al. [6]. The Al flyer in the experiments is a cylindrical disc and is modeled here in 2D as a rectangle of thickness 25 μm that longitudinally extends out of the computational domain. The crystal and binder level sets are created using nano-CT images[125]. The image processing methods adopted to obtain levelset fields from the images have been explained in [126]. Particular to this work, the binder levelset is created by performing a Boolean operation that 'subtracts' a crystal level set (without heterogeneities such as cracks/voids) out of the level set defining a rectangular block of binder material. The crystal level set, containing heterogeneities, is then placed back in the 'cutout' to complete the crystal-binder aggregate system. The above process is depicted in Fig. 1(a). Any region not occupied by material (HMX/binder/Al) is treated as void.

To compare with the explicit modeling of the Al flyer striking the block of binder and its contents, we also study the treatment in which the impact is modeled using a shock-pulse boundary condition. In this case, the computational domain contains only the crystal and binder levelsets, and replicates the setup reported in Roy et al. [1]. All domain boundaries, except the one with the shock-pulse boundary condition, are imposed with Neumann (zero-gradient) condition for all flow properties.

The domain size for the cases reported in Sec 3.1 is 80 μm × 120 μm for the shock-pulse boundary condition case and 80 μm × 150 μm for the flyer impact case. The latter domain size is retained for the calculations given in Sec 3.2-3.3. The high-resolution calculations shown in Sec 3.4 are conducted for larger samples with domain size of 300 μm × 330 μm. The domains are discretized spatially with a uniform spacing of 80 nm. All calculations were conducted on a HPE Cray EX system which has compute nodes with 2.6GHz AMD 7H12 Rome 128 core processors and 238GB of memory. Depending on the computational load defined by the number of grid cells, which varied across the cases from 1.9M cells to 15.3M cells, the simulations had resource usage from 132 cores and 1000 CPU hours to 1760 cores and 102,080 CPU hours. The computational requirements for the high-resolution cases reported in Sec 3.5 were higher and are mentioned separately in that section.

## 2.3 Data diagnostics and comparison to experimental results

The details of the experimental procedures and diagnostics are given in Roy et al. [1]. Briefly, the shock initiation experiments were conducted through laser-launched flyers impacting a sample of deconstructed PBX consisting of a binder-encased single HMX grain. High-resolution x-ray computed tomography (CT) scans of the PBX samples (taken before the experiment) were used for the tandem computations. Quantitative estimation of hotspot thermal output was performed using high temporal resolution optical pyrometry.

Since spectral radiance-based pyrometry data obtained in the experiments yielded spatially-averaged values inherently biased towards the peak temperatures[6], in order to have a faithful comparison with experiments the simulated temperature field (at a given timestep) was cast into a histogram distribution and the area-weighted spectral radiance for each histogram bin was obtained using Planck's radiation law, as done previously in Roy et al. [1]. The temperature (alongside emissivity) was then obtained by fitting a curve to the wavelength-radiance distribution.

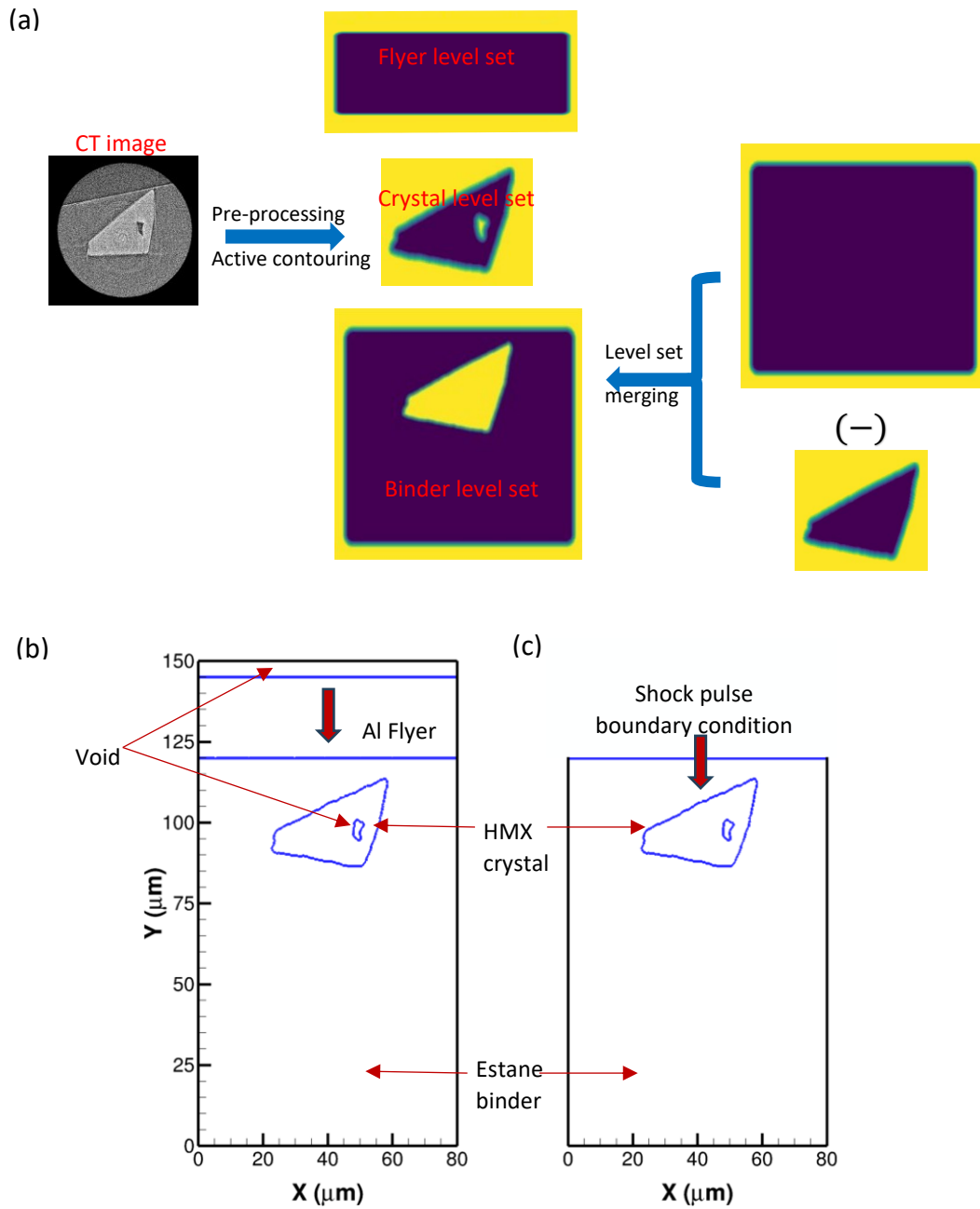

*Figure 1. Computational setup. (a) Procedure for creating level sets of flyer, binder and crystal. (b) Setup for simulation of a flyer impacting a crystal embedded in binder. (c) Setup for simulation of a crystal-binder system shocked by an impact modeled using a shock-pulse boundary condition at the top boundary.*

## 3. Results and Discussion

In the following sections, we examine the different aspects of meso-scale numerical simulations that bear upon physically accurate representation of the impact-induced initiation of a PBX material. The systematic study below establishes the relative importance of various treatments of the material boundaries – both external and internal (i.e., interfaces) – and the models that are used to represent the thermo-mechanics of the material.

### 3.1 Effect of boundary conditions: explicitly tracking flyer impact vs modeling impact through a shock-pulse boundary condition

In this section, we analyze the differences between the flow solutions yielded by two approaches for simulating the impact of an Al flyer on a crystal-binder aggregate: A) explicitly tracking the motion of the flyer as it hits the binder surface, and B) modeling the impact-induced shock via a velocity pulse applied as a domain boundary condition to the binder surface. The calculations for both the setups in this case are conducted in inert mode, i.e., by turning off the reaction chemistry model, to focus on the thermomechanical aspects.

Figure 1 (b & c) illustrates the two setups. In Fig. 2, the numerical schlieren image of the transient flow solution from the two setups is tracked in the first two columns, while the third column tracks the centerline pressure values. In setup A (left column in Fig. 2), the Al flyer impacts the binder at a velocity of 3 km/s at t=0. The elastoplastic deformation at the interface between the flyer and binder generates shock waves (in both media) which are captured using the high-order scheme. In setup B (middle column in Fig. 2), the impact from the flyer on the binder is modeled via a hybrid Dirichlet-Neumann condition imposed at the top boundary. The binder surface is supplied with a particle velocity ($U_P$) in the form of a square pulse signal – a 'shock rise' consisting of an initial $U_P$ of 2118 m/s imposed starting at t=0, sustained for a duration of 5.5 ns, and a 'shock decay' modeled by setting $U_P$ to 0 m/s for 20 simulation timesteps. Thereafter the boundary is switched to a zero-gradient (Neumann) type for the remainder of the simulation. The initial $U_P$ and shock plateau intervals were determined from the calculation using setup A and was also cross-checked to match with the PDV measured velocity value for a 3 km/s flyer impact experiment on a polymer-embedded HMX crystal reported in Johnson et al.[127]. This cross-check incidentally confirms that the Riemann-based interface treatment accurately captures the impedance mismatch between the flyer and the binder to yield the correct $U_P$ at the binder surface.

The impact between the flyer and binder generates a plastic shock in each of the materials. In Fig. 2(a), the flow-field at 3 ns after the impact is seen, with the pressure immediately downstream of the shock matching closely across the two setups. In setup A, the shock reflected off the HMX-binder interface transmits into the flyer, while in setup B it is reflected into the domain as spurious waves owing to the Dirichlet condition applied at the boundary at 3 ns. As seen in the centerline pressure profiles in Fig. 2 (right column), the pressure profile in setup B is affected by these waves, albeit only to a mild degree at the shock front. Due to the absence of a flyer material boundary in setup B, the movement of the binder interface is not actually tracked and only the region just above the crystal is physically consistent; the rest of the material is an unphysical extension of material entrained into the domain from the boundary. This situation is typical of boundary-condition-based models seen in literature [11, 83]. Explicit modeling of flyer impact as in Setup A obviates such problems arising from approximation of boundary conditions at the expense of complicating the computational setup. We further examine the implications of these two approaches on the actual pore collapse result in the following.

In Fig. 2(b), the prominent features seen at 5 ns are the collapse of the pore in the crystal and the shockwave in the flyer being reflected from the opposite (free) end as a relief wave. The pressure profiles at this stage display minor deviations across the two setups. At 9 ns, in Fig. 2(c), the pressure plot shows how the relief wave transmitted into the binder in setup A expands the previously compressed materials – the mechanics of this are well emulated by the shock sustenance and decay phases in setup B. The instances of 14 ns and

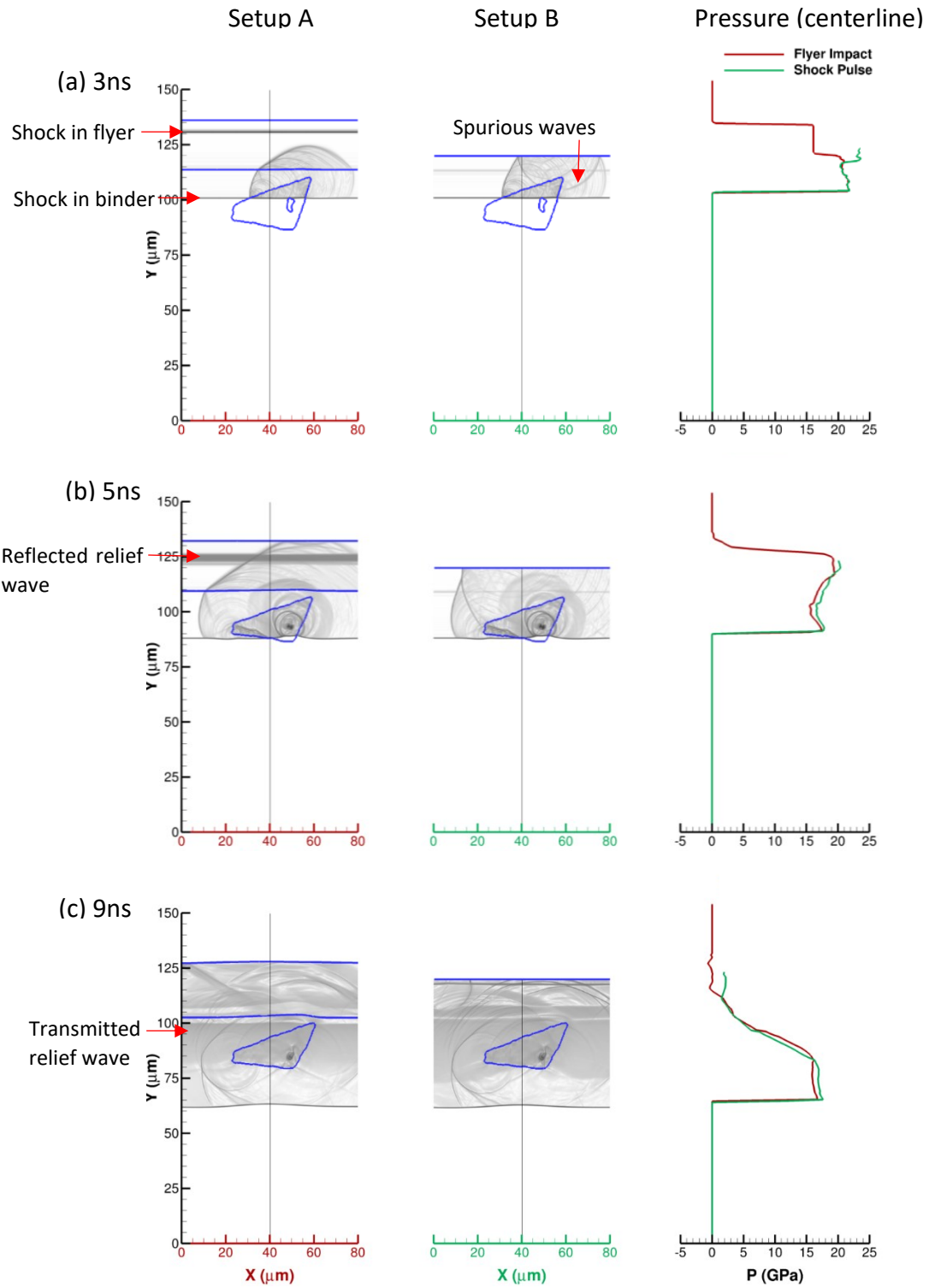

Figure 2. Schlieren rendition of density contour plots from simulation of crystal-binder system shocked via (left) flyer impact, and (center) shock pulse boundary condition. (Right) Centerline pressure plot comparison for the two cases. (a) 3ns. (b) 5ns, (c) 9ns.

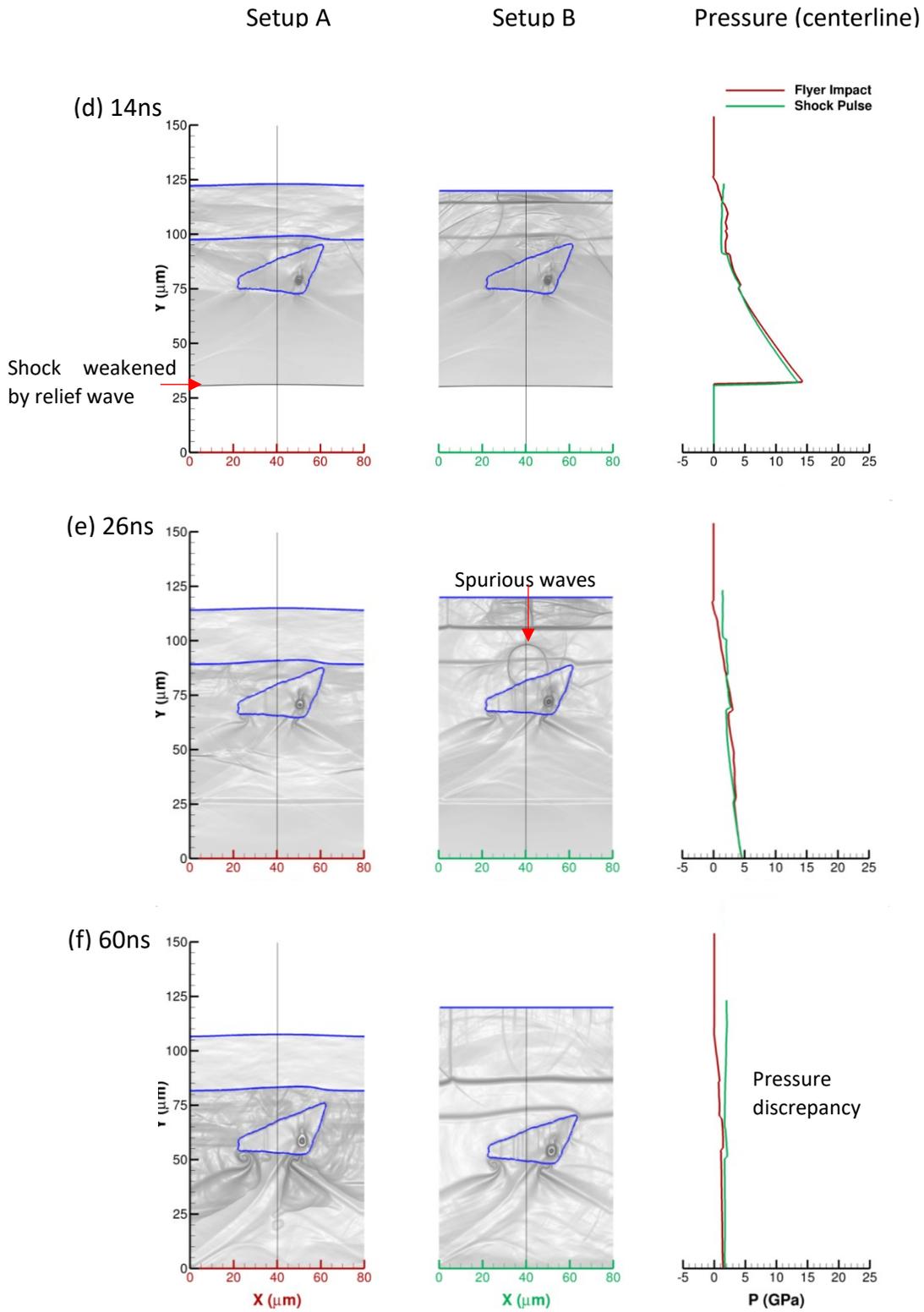

Figure 2. (contd.) Schlieren rendition of density contour plots from simulation of crystal-binder system shocked via (left) flyer impact, and (center) shock pulse boundary condition. (Right) Centerline pressure plot comparison for the two cases. (d) 14ns. (e) 26ns, (f) 60ns.

26 ns (Fig. 2 (d, e)) both show spurious waves in setup B but with minor perturbation to the pressure profiles. However, at 60 ns (Fig. 2(f)), discrepancies between the pressure profiles begin to show. In setup A, the pressure approaches ambient conditions at the top material boundary, whereas in setup B the Neumann boundary condition prevents the pressure from matching ambient values. This leads to a pressure mismatch between the two cases towards the top portions of the binder as seen in the figure.

The analysis presented here shows that if the particle velocity at the binder surface $U_P$ and shock-pulse interval are accurately known, the approach of using a shock-pulse boundary condition as in setup B can be a simple yet reasonably physical approximation to modeling of the flyer impact that captures the initial shock response. However, it is to be noted that only a carefully tuned pulse-interval of 5.5 ns (approximated using Eq. (19) in [41] and then fine-tuned by trial-and-error), could reproduce this effect specifically for the interaction between the Estane binder and the 25 μm thick Al flyer. For each flyer thickness and velocity, the corresponding pulse width must be similarly determined. Even when this is done, there are noticeable discrepancies in the flow physics between the two approaches, particularly once the shock has passed and the long-time scale response becomes of interest. Flow features near the impact location at longer duration can become unreliable, as a discrepancy in pressure develops. The centerline pressure in setup B in this case of a 3 km/s impact was approximately 1.2 GPa higher than the pressure at the binder surface in setup A. This difference, although not large, can become significant in the reactive case, such as in affecting the growth and intensity of smoldering hotspots in the wake of a shock passage – an aspect revisited in greater depth in the next section.

In summary, simulating the multi-material impact in lieu of approximating the impact through a boundary condition is a more direct approach that yields accurate thermo-mechanical response of the shocked PBX. Wave transmissions and reflections are naturally accounted for, which prevents the emergence of spurious waves and captures a physically consistent post-shock compression state. The advantage of the boundary-condition-based approach, namely its simplicity, breaks down when the profile of the pulse cannot be easily determined. Roy et al. [1] used a 4 ns pulse to approximate the impact by a 25 μm flyer, informed by the measurements for a ~3.5 km/s impact reported in [6], but the accurate duration to replicate a 3.3 km/s flyer impact was determined, by trial-and-error, to be 5.3 ns for use in this work. Particle doppler velocimetry (PDV) measurements typically have ~1 ns temporal resolution[128, 129]. The inherent resolution of the PDV system, signal rise times, and signal quality (e.g., large drops in signal at impact), can smear a significant number of interference fringes making it difficult to resolve the exact pulse width, particularly in this strong shock regime. The pulse width can have a strong influence on the pressure profile and hence on the shock sensitivity. For a given flyer velocity, the effect of different pulse durations on the shock initiation of a crystal using setup B has been demonstrated in Fig. S1 in Supplementary Information, showing how sensitive the hotspot initiation and growth can be to slight variations of pulse durations in the boundary-condition – based approach. Further, the shock can be non-planar (2D) if the flyer is not a regularly shaped cylinder which is wide enough to cover the impacting surface, generating corner effects. Given a framework complete with reliable material models (calibrated either empirically or derived from first principles), accurate shock-capturing schemes, and physically consistent interface treatments, physically accurate boundary conditions are best applied by tracking a flyer in the simulation. Explicit flyer tracking is therefore adopted in all calculations to follow in this paper.

## 3.2 Effects of HMX strength models at low (1km/s) velocity flyer impact

Next, we examine the effects of the two strength models for HMX – EPP (constant yield stress) and M-JC (rate-dependent yield stress) – on the stresses and deformations in the bulk and at interfaces of the shocked material. Higher shock pressures are likely to yield hydrodynamic behavior for both the models due to pronounced thermal softening effects and/or the loss of strength due to melting. Hence, a low shock strength scenario via a 1 km/s flyer impact has been utilized to exemplify the differences in pore collapse dynamics and hotspot formation arising from the strength model for the HMX crystal.

The low velocity flyer impact (at 1 km/s) creates a moderate shock with particle velocity of 550m/s and ~ 4 GPa of pressure in the HMX crystal. Figure 3 shows the temperature profiles at two timestamps, 7 ns and 8 ns, capturing the collapse of the single pore as the shock passes through the crystal. The EPP model (Fig. 3(a)) displays the signature of inertia-dominated collapse, even at the low shock pressure. The stresses around the pore are high enough to exceed the constant yield threshold of 260MPa, readily causing plastic deformation. The material at the leading tip of the pore to the shock (at 7ns) experiences large strains and is seen jetting into the opposite end in longitudinal direction to the shock (at 8ns). The M-JC model (Fig. 3(b)). shows a distinctly different mode of collapse. The collapse process is shear-dominated [15, 130], and the energy in the shock is dissipated via shear bands. The pore collapses in a transverse direction to the shock. This "pinching" type collapse inhibits material jetting. In recent works, molecular dynamics studies of void collapse in HMX have shown the existence of extensive network of shear bands in such pore collapse events [56, 58]. Thus, the atomistics-consistent M-JC model is physically more consistent compared to the EPP model in capturing the deviatoric stress effects at low shock strengths.

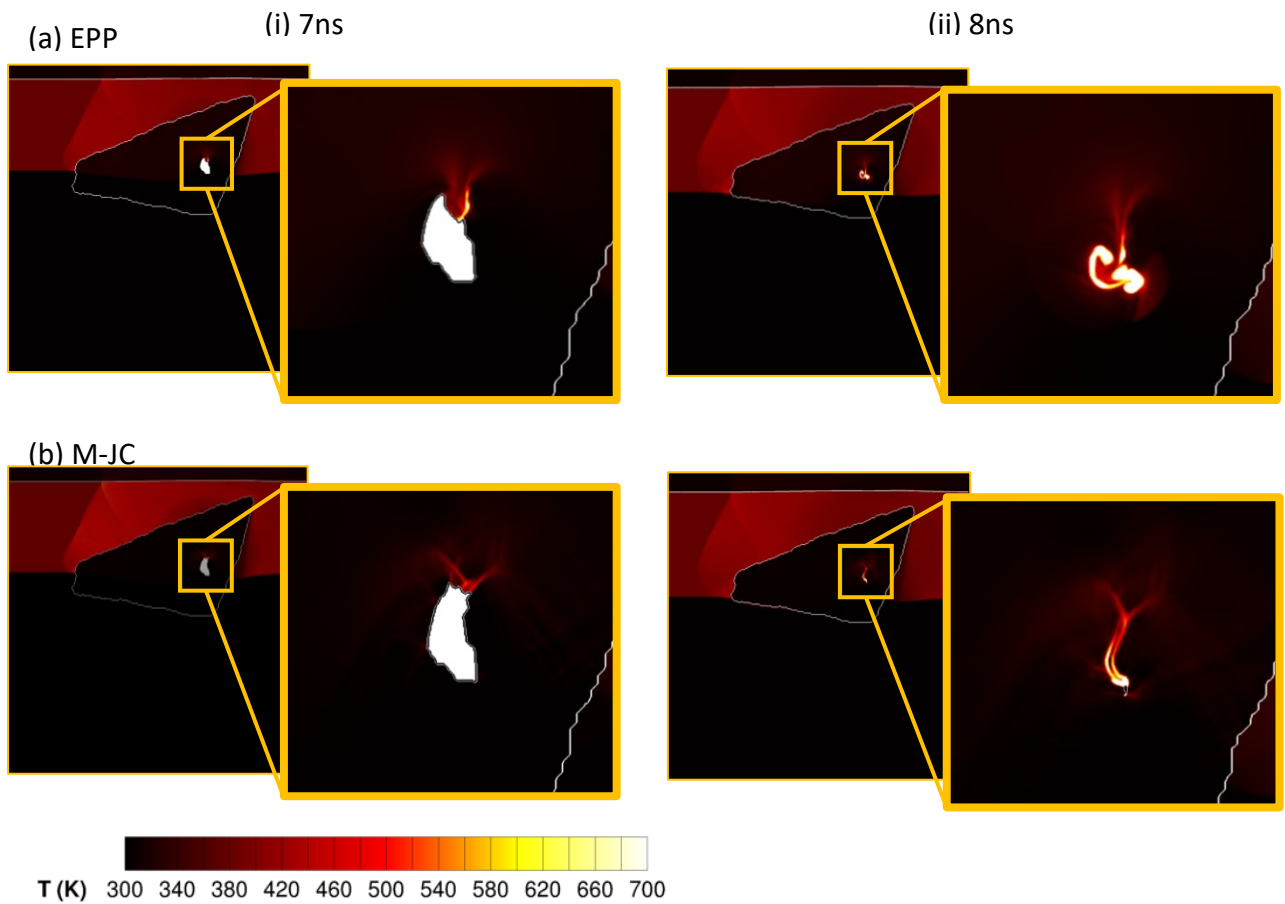

*Figure 3. Temperature contour plots of simulated 1km/s flyer-impact on a crystal-binder aggregate showing the differences in the mechanism of pore collapse in response to the overpassing shock for two HMX material models at (i) 7ns and (ii) 8ns. For each snapshot, the inset shows a zoomed view on the right. (a) Constant yield stress model (EPP), 7ns and 8ns respectively. (b) Johnson-Cook rate-dependent yield stress model (M-JC), 7ns and 8ns respectively.*

## 3.3 Effects of HMX material models on energy localization for high velocity (3.3km/s) flyer impact

We now systematically analyze the effect of material models for HMX under the high velocity impact conditions simulated in Roy et al. [1], corresponding to the experiments of Johnson et al. [6]. Table 1 summarizes the two modeling aspects which most significantly impact the energy localization as shown in Lee et al. [19], namely the specific heat model and the material strength model. Thus, we examine: 1) effect of treatment of the specific heat: constant value of 2359 KJ/kg as in classical MD calculations vs a temperature-dependent model (see equation S20 in Supplementary Information), and 2) effect of the plasticity model for HMX: an EPP model based on constant yield-stress of 260 MPa [47] vs the rate-dependent Johnson-Cook model (M-JC) [58]. The comparative study investigates the effects on the flow field and hotspot characteristics due to the changes in the different model configurations listed in Table 1.

Table 1. Case matrix with details of model configurations used for a 3.3kmps flyer impact calculation.

| CASE | SPECIFIC HEAT ($C_v$) MODEL | HMX MATERIAL MODEL |
|---|---|---|
| 1 (baseline) | Constant $C_v$ (classical) | Constant yield stress |
| 2 | ***Temperature dependent*** | Constant yield stress |
| 3 | Temperature dependent | ***Rate-dependent yield stress (M-JC)*** |

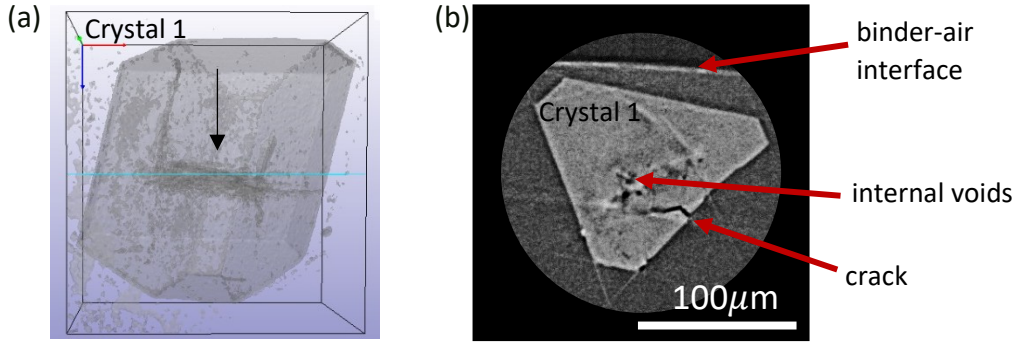

*Figure 4. (a) Three-dimensional rendering of a polymer-encased HMX crystal. (b) Selected cross-sectional CT image of the crystal used for simulations. The crystal has large internal voids and cracks. The green line indicates the precise location from which the n-CT cross section was extracted. The black arrow shows the direction we look at the cross-sectional CT image. This figure has been reproduced from Roy et al. 2022 [1].*

Results are obtained using the hierarchy of model implementations in Table 1, for a 3.3km/s flyer impact calculation on an imaged geometry of a sample of HMX crystal embedded in binder, as shown in Fig. 4. The sample analyzed is one of the three imaged crystals previously studied in Roy et al. [1]. The cross-section was selected to have distinctive features of the heterogeneities in the crystal. The crystal is characterized by a cluster of internal pores, sharp corners, and an extended crack running from the center to one of the edges. A strong shock at the high impact velocity of 3.3 km/s induces ~20 GPa of pressure in the binder and ~25 GPa of pressure in the HMX cystal. This pressure regime has been known to readily produce and sustain hotspots in damaged crystals[1, 6], and is a good test for observing the effects of various modeling techniques on the mechano-chemical aspects of hotspot criticality. Contour plots of pressure and temperature at two instances for the shocked PBX sample are compared in Fig. 5. The panels in the left column show the pressure field and the panels in the right column show the temperature field. The first instance (i & iii) corresponds to t = 13.5 ns showing conditions in the sample immediately after the shock passes over the defects, localizing energy at the hotspots. The second instance, labeled (ii & iv) shows the state at a later time (t=40ns) long after the initiation event due to the shock. The rows correspond to the different model changes indicated in Table 1.

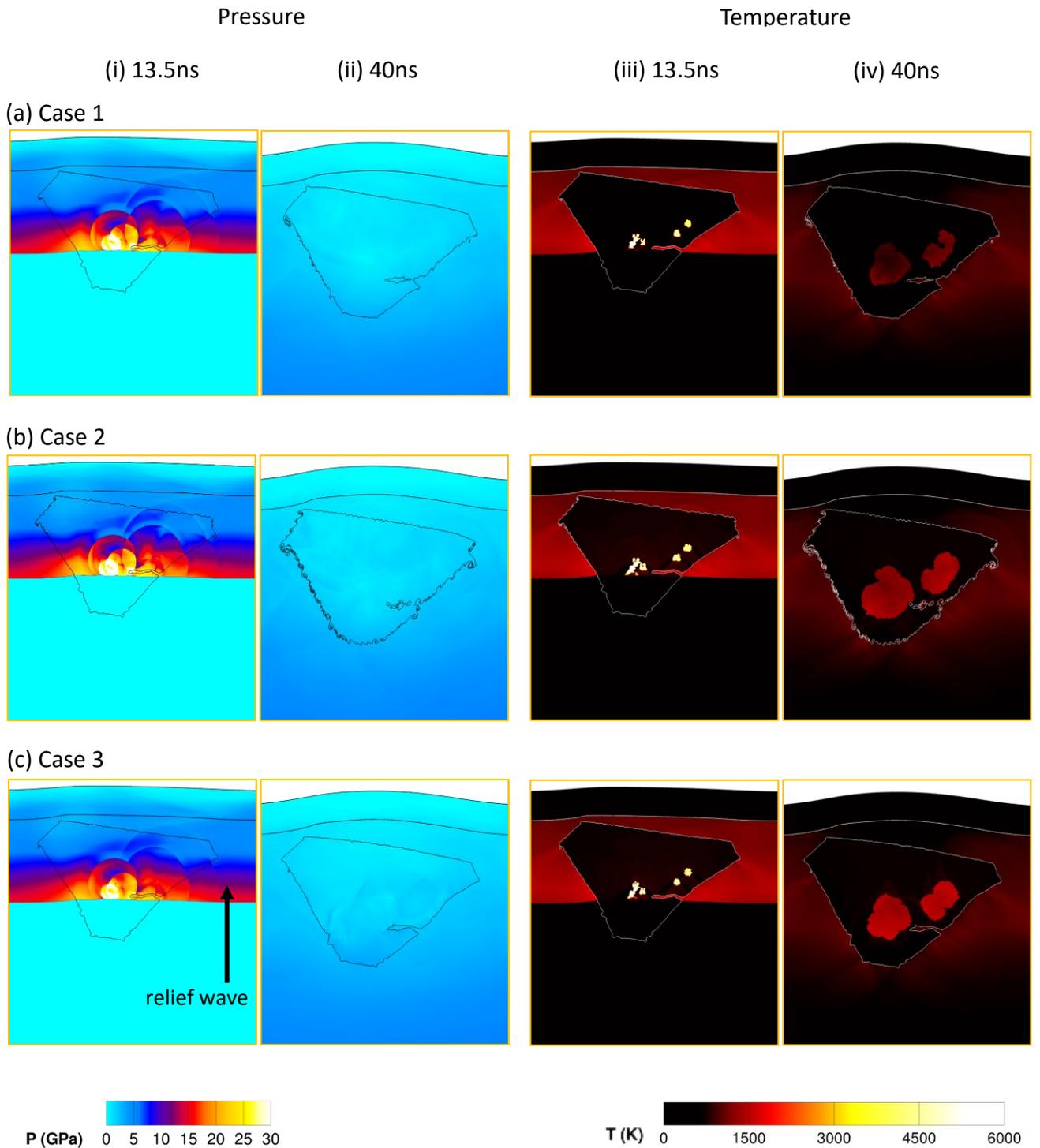

Figure 5. Pressure and temperature contour plots from simulation of a 3.3km/s flyer impact on the PBX sample shown in Fig. 4 using different setup techniques and models given in Table 1. (a) Case 1: Constant $C_v$, constant yield-stress model (EPP). (b) Case 2: Temperature dependent $C_v$, constant yield-stress model (EPP). (c) Case 3: Temperature dependent $C_v$, rate-dependent yield-stress model (M-JC).

Going from the baseline configuration of Case 1 with a constant (classical) value of 2359 KJ/kg for specific heat ($C_v$) in row (a), to that in Case 2 with temperature-dependent relation for $C_v$ in row (b), the thermal profiles at 13.5ns are mostly identical across the two cases since the temperature dependent $C_v$ asymptotes to the classical value at the high temperatures prevalent at this time instant (Fig. 2(c) in [56]). However, the hotspots cool down at 40ns and the values of $C_v$ from the two models differ significantly. The temperature-dependent value of $C_v$ is lower than the constant value in the classical treatment at 40ns, influencing the temperature fields differently at this instance. This is evident since at 40 ns the hotspots in Case 2 appear to be hotter (brighter) and have grown larger than Case 1. Next, the rate dependent yield-stress model (M-JC) is added in Case 3 shown in row (c), switching out the EPP model of Case 2. The plots show a number of differences between both instances of time. At 13.5ns, the void collapse events in Case 3 (Fig. 5(c-i)) are less violent as visualized from the more smeared appearance of the weaker blast wave boundaries, when compared to Case 2 (Fig. 5(b-i)) showing wave interactions of sharper contrast. Further, at 40ns, the crystal boundary in Case 3 (Fig. 5(c-i)) is smoother than for the other cases, missing the high levels of Richtmyer-Meshkov[131, 132] instability with vortical structures seen all along the interfaces in Case 2. The above differences are explained by the nearly hydrodynamic behavior of the EPP model when compared with the rate-dependent hardening effects of the M-JC model. In the latter model, as in Case 3, the strain and strain rate hardening in the JC model enhances the strength of the crystal and suppresses the interfacial instabilities.

For a more quantitative comparison of Cases 1-3, the spatially-averaged temperature is evaluated from the simulated data at intervals of 1 ns using the procedure discussed in Sec. 2.3., and compared across the current set of calculations; but we also include the simulated and experimental data from Roy et al. [1]. The plotted values are shown in Fig. 6. The spectral-radiance based temperature evaluation is biased towards the higher temperatures in the dataset. Therefore, this methodology tracks the highest temperature excursions in the sample as a function of time after the impact and the approximate interval of temperature spike during the initiation event; the one-to-one comparison of instances of temperature rise is not feasible since the computations are only done on a representative 2D cross-section of the three-dimensional crystal in the actual experiments which can have thermal events out-of-plane from the one simulated. Despite this, in Fig. 6, it is seen that the present models give peak temperatures which are in much better agreement with experiments (5800 K - 6700 K) when compared with the under-predicted values in Roy et al. [1] simulation. The Roy et al. [1] data shows a peak at 5ns due to the surface reaction which starts at the incidence of shock on the crystal surface (see Fig. 4 in Roy et al. [1]). As noted earlier, surface reaction is not present in the calculations with the updated models. Whether the earlier peaks in the experiments (at ~5ns) are due to surface reaction or due to the collapse of pores outside of the simulated cross-section (e.g. ~5ns and 25ns), cannot be ascertained. Among cases 1-3, the peak temperature in Case 1 reaches 8300 K, while that in Case 2 reaches over 7900 K. Model 3, with a peak of 6600 K is the closest match to the experimentally recorded thermal spike during the initial pore collapse and hotspot generation event at 8ns. Beyond this instance of time, we see the simulated temperatures from all the models drop faster than in the experiments, though the general duration of high thermal signature extends to approximately 25 ns in both simulations and experiments, at which point the crystal is nearly fully consumed by the deflagration process.

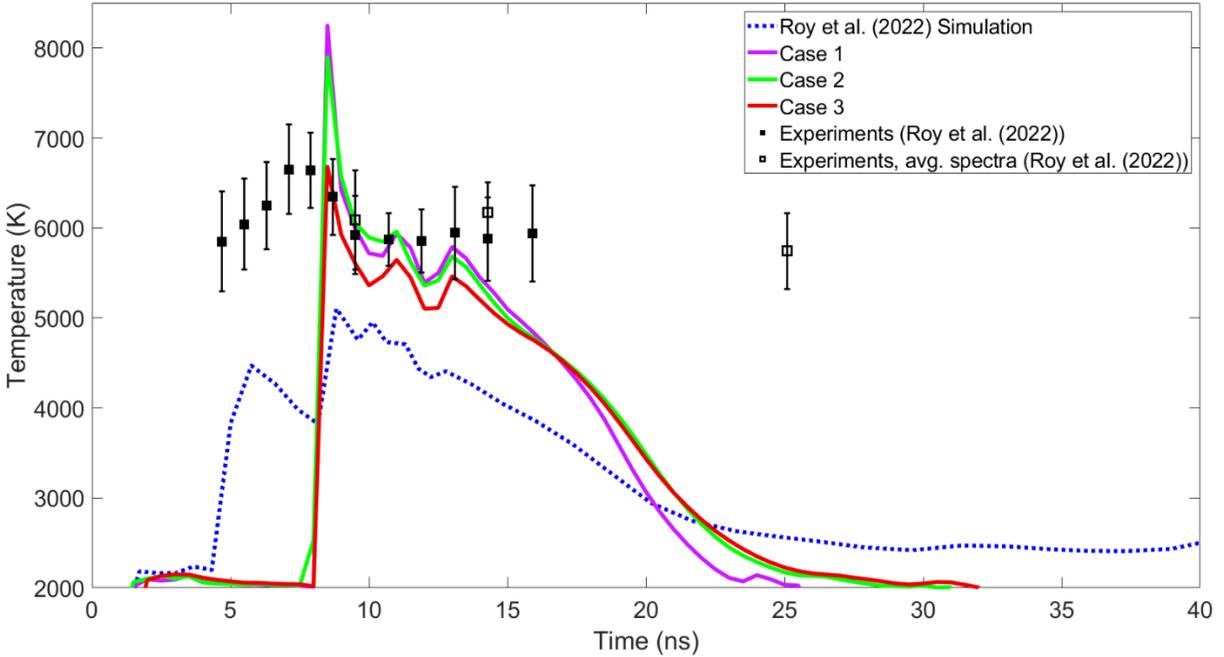

*Figure 6. Time-dependent (spatially averaged) temperatures from simulation of a 3.3km/s flyer impact on the PBX sample shown in Fig. 4: comparison of data from experiments and simulation using past models (Roy et al. [1]), and present computational models and setups. (cases 1-3 in Table 1).*

In summary, we infer that within the window of the highest temperature events of 5 ns – 15 ns, the temperatures predicted by the highest fidelity model configuration (Case 3) have the best agreement with the experiments and show improvement over those predicted by the models reported in the Roy et al. [1] study. In this time span of 5 ns – 15 ns the hotspots are due mainly to the collapse of pores and other internal defects in the wake of the overpassing shock. While the new models have brought the peak temperatures to better agreement with experiments, there are yet considerable challenges to a head-to-head comparison of hotspot temperature evolution in the short time scales of shock initiation in both experiments and continuum calculations, as noted also by the authors in Roy et al. [1]. Explanations for some of the discrepancies between simulations and experiments may be difficult to provide until high-resolution 3D DNS of such PBX samples become computationally feasible. Rough estimations indicate that 3D calculations will entail grids amounting to tens of billions of grid points if shock initiation calculations were to be performed on these PBX samples maintaining the same level of fidelity through grid resolution. Clearly such calculations will be challenging for several more years given current computational resource structures.

### 3.3.1 Pore collapse, reaction and hotspot growth in a single shocked HMX grain:

Having gained an understanding of the impact of the numerical and modeling updates on energy localization at hotspots, we revisit the results from the final model configuration of Case 3 in Table 1 to study the shock sensitivity of the PBX sample shown in Fig. 4 and to contrast it with previous calculations [1]. The simulation movies are presented in Supplementary Information. The calculation is conducted with the rate-dependent yield-stress (M-JC) and temperature dependent $C_v$ for HMX. The Al flyer is imparted an initial velocity of 3.3 km/s, corresponding to particle velocity of $U_P$ of 2.3 kmps at the binder surface; the impact between the flyer and the binder occurs at t=0. In addition to the pressure and temperature plots in Fig. 5(c), Fig. 7 shows contour plots at four selected instances of time; row (a) shows the numerical schlieren rendition of density gradients highlighting the shock dynamics, while the plots of mass-fraction of the final reacted

HMX species ($Y_4$) are given in row (b) to track the extent of reaction via the burn fronts. The snapshots selectively show, in sequence, the instance when the shock just hits the top portion of the crystal (5.5 ns), the instance when the pores and defects are just collapsed by the shock (13.5 ns), the instance when the relief wave catches up to the shock (18 ns), and the state at a later time after the shock passage (40 ns). Additional plots can be found in Figs. S3 & S4 in Supplementary Information, where the contour plots of pressure and temperature at the four instances for both the crystals are provided.

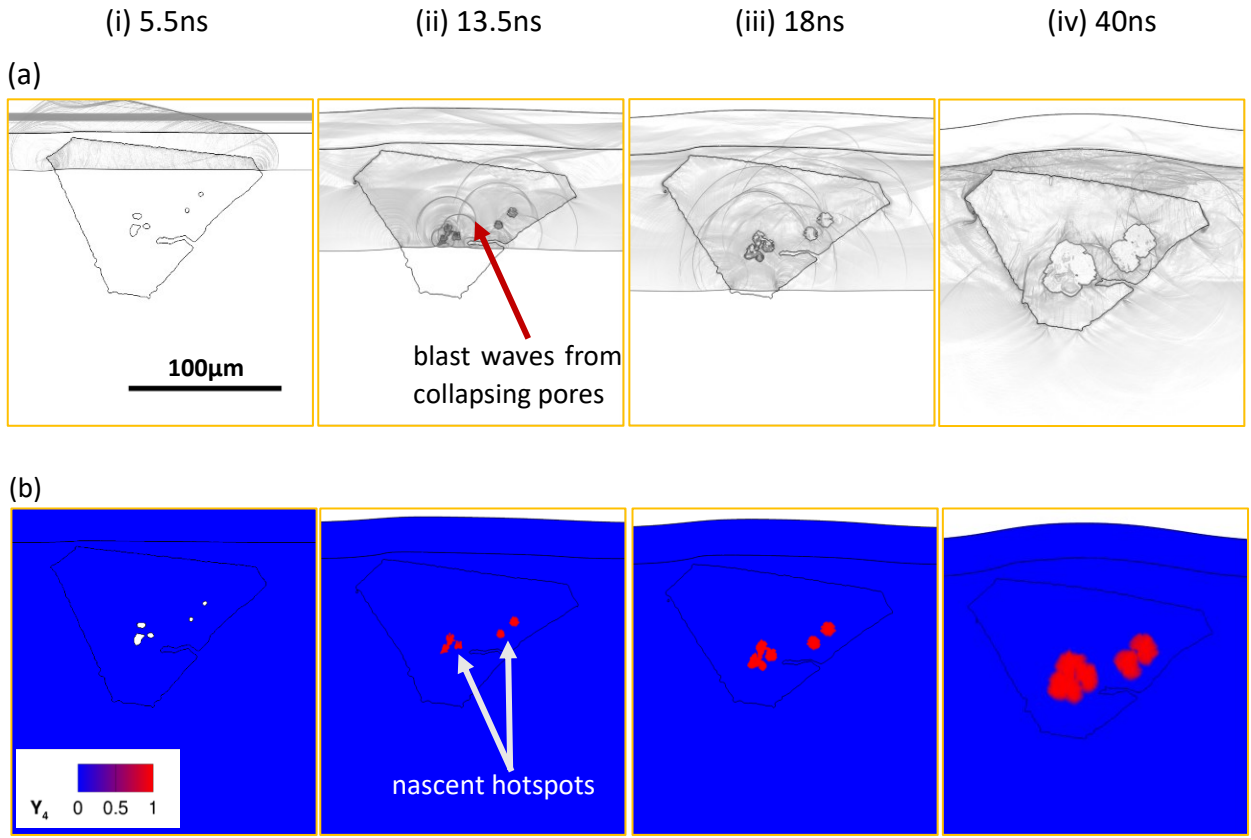

*Figure 7. Plots of (a) numerical schlieren and (b) mass-fraction of reacted species, from simulations of a 3.3km/s flyer impact on a HMX crystal encased in binder. The impact from the Al flyer is explicitly tracked. Rate-dependent yield-stress model (M-JC) and temperature-dependent Cv is used for HMX.*

At 5.5 ns, the shock in the homogeneous part of the HMX crystal without any pores or cracks creates only modest temperature rise without reaction (Fig. 7(b-i)). At 13.5 ns, the collapse of several internal pores is observed in Fig. 7(a-ii), creating nascent hotspots. Blast waves emanate from the sites of collapse, seen in the schlieren, and local temperature at the hotspots reach upwards of 6500 K (Fig. 5(c-iii) & Fig. 6). Unlike the pores, it is observed that the crack on the right edge does not collapse due to the binder material within it. At 18 ns, the reaction from the hotspots has propagated as seen in the $Y_4$ plots in Fig. 7(b-iii). The pressure plot in Fig. 5(c-i) shows the relief wave passing over the shocked crystal just before catching up to the leading shock. Lastly, the state of the reaction at 40 ns after the impact is shown in Fig. 7(b-iv), with the pressure reaching near ambient values (Fig. 5(c-ii)) and the hotspots cooled to below 2000 K (Fig. 5(c-iv)).

Besides the peak temperature, the simulated fields reported here have a noticeable difference from the ones reported in Roy et al. [1]. In Roy et al. [1] all the crystals showed extensive reaction at the surfaces in the wake of the shock which is absent in the present case (Fig. 7(b)). This resulted in a higher amount of reacted mass in the crystals in [1]. The above difference can be attributed to the cumulative effect of the following changes in the numerical aspects between the two sets of calculations. First, the calculations in Roy et al. [1] used a 4 ns shock pulse for $U_P$ 2.3 km/s instead of tracking the movement of the Al flyer as in the present case. The residual pressure in the domain when employing a shock-pulse boundary condition, discussed in Sec. 3.1, can affect the hotspot growth. Second, the calculations in Roy et al. [1] used a previous version of the rate-dependent Johnson Cook model for HMX compared to the M-JC model used here which additionally accounts for softening effects at high strain rates (occurring beyond a threshold stress limit but well below hydrodynamic conditions). Third, the calculations in Roy et al. [1] were conducted using $3^{rd}$ order accurate ENO scheme which is more dissipative than the $5^{th}$ order WENO scheme in the present calculations. The shift in the spatially-averaged hotspot temperature profiles due to the collective effects of the above differences in interfacial treatments, material models and numerics can also be seen in Fig. 6.

### 3.4 High-resolution calculation of a shocked crystal-binder system

The analysis in the preceding sections demonstrated the suitability of the models (as listed for Case 3 in Table 1) for capturing shock-induced deformations and hotspot generation and growth in HMX based PBXs. With these upgraded implementations, we push the envelope on fidelity by conducting an ultra-high-resolution DNS calculation as a demonstration of the capabilities of the improved models and numerics. The case chosen is that of the 1km/s impact of an Al flyer on a PBX sample, earlier depicted in Sec. 3.2. The low shock strength has been chosen to bring out the shear effects in the crystal, which are otherwise overwhelmed by hydrodynamic pressure at higher impact velocities. However, in the present case the solution is recalculated with significantly increased resolution – the grid size is 5nm, approaching atomistic scales, for an overall domain of size 50μm×70μm. The uniformly spaced mesh comprised of about 140M grid cells. This highly resolved computation ran on 7000 CPU cores for ~400 hours of wall-time, i.e., about 2.8M CPU hrs.

The numerical schlieren contour plots in Fig. 8 demonstrate the efficacy of the high-order schemes and high-resolution case setup by revealing the rich wave dynamics and propagation of shear bands. Multiple small 'wavelets' are visible emanating from the undulations of the rough crystal boundary as the shock impinges on the top surface of the crystal. Reflected shocks, blast waves, relief waves, and transmitted waves are vividly captured in the 10ns span of simulation. Further, a dense network of needle-like density gradients in the shocked crystal is also observed. The majority of these sharply resolved striations are also captured in the contours of temperature and von-Misses strain in Fig. 9, which suggest that these are shear-localized hotspots since density gradients are accompanied by temperature and strain upticks that are well above the bulk values of the shocked HMX. The low dissipation $5^{th}$ order WENO-Z scheme preserves the resolution of the bands on the fine 5nm mesh. Here, shear bands are seen emanating from the crystal surface impinged by the incident and reflected shocks in addition to within the stressed material around the collapsing pore. To the best of the authors' knowledge, this is the first visualization of shear bands emanating from the surface of a real (imaged) HMX crystal geometry. The interface treatment owing to the HLLC Riemann based GFM [69, 70] plays an important role in accurately computing the intricate waves and disturbances from disparate material interfaces (crystal-binder, crystal-void, etc.) that eventually enables the M-JC strength model to realize these elastoplastic flow features. Such surface shear bands have previously been reported by Springer et al. [15], albeit from idealized semi-circular pores at the surface of a block of HMX. The temperature rise of 500 K in the shear bands and 700 K due to pore collapse agrees well with values reported in the literature [58, 118].

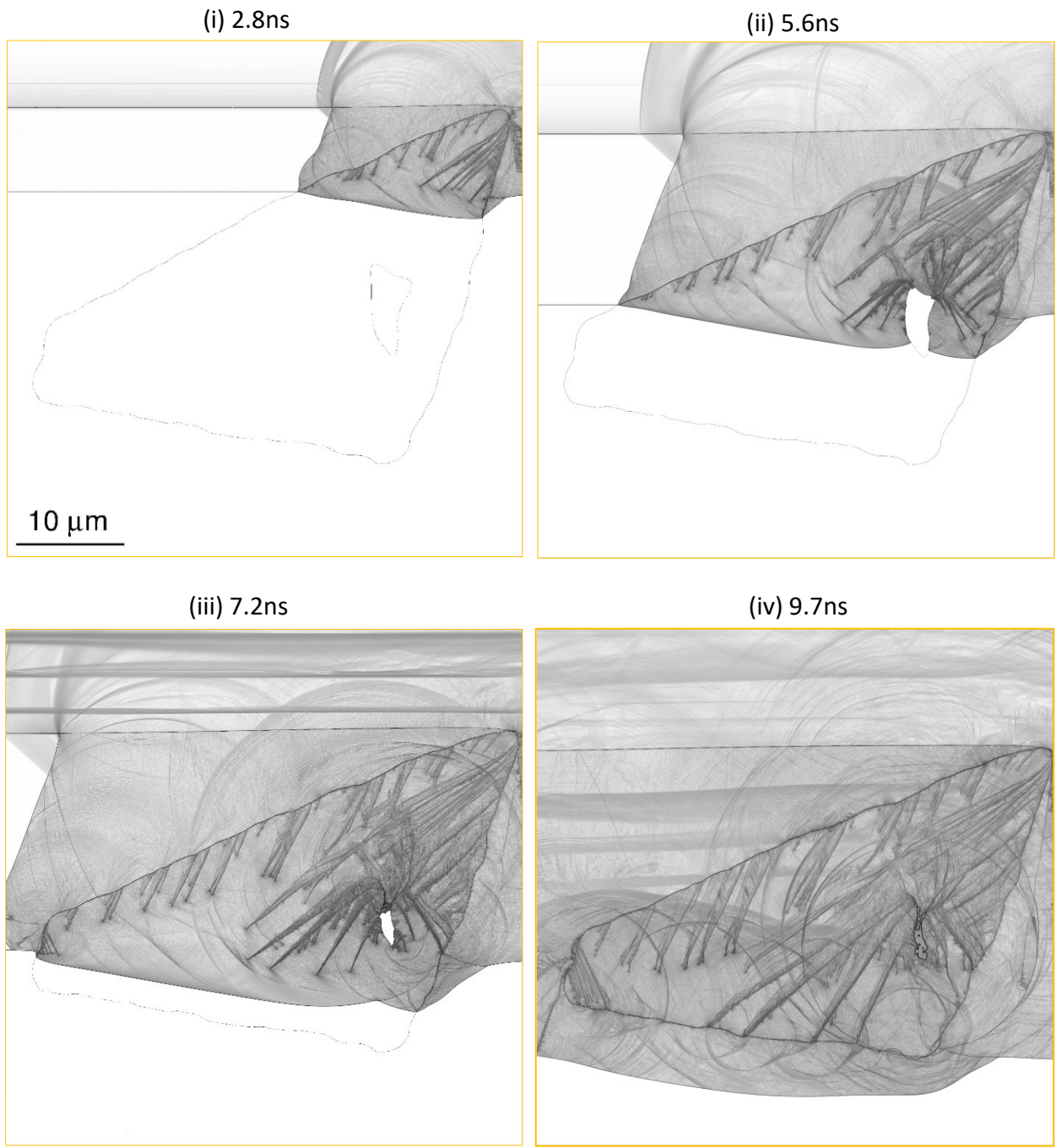

*Figure 8. Numerical schlieren plots from high resolution simulation of a 1km/s flyer impact on an HMX crystal embedded in binder at (i) 2.8ns, (ii) 5.6ns, (iii) 7.2ns, and (iv) 9.7ns. The impact from the Al flyer is explicitly tracked. Rate-dependent yield-stress model (M-JC) and temperature-dependent $C_v$ is used for HMX. 5$^{th}$-order WENO with Riemann-based GFM has been used.*

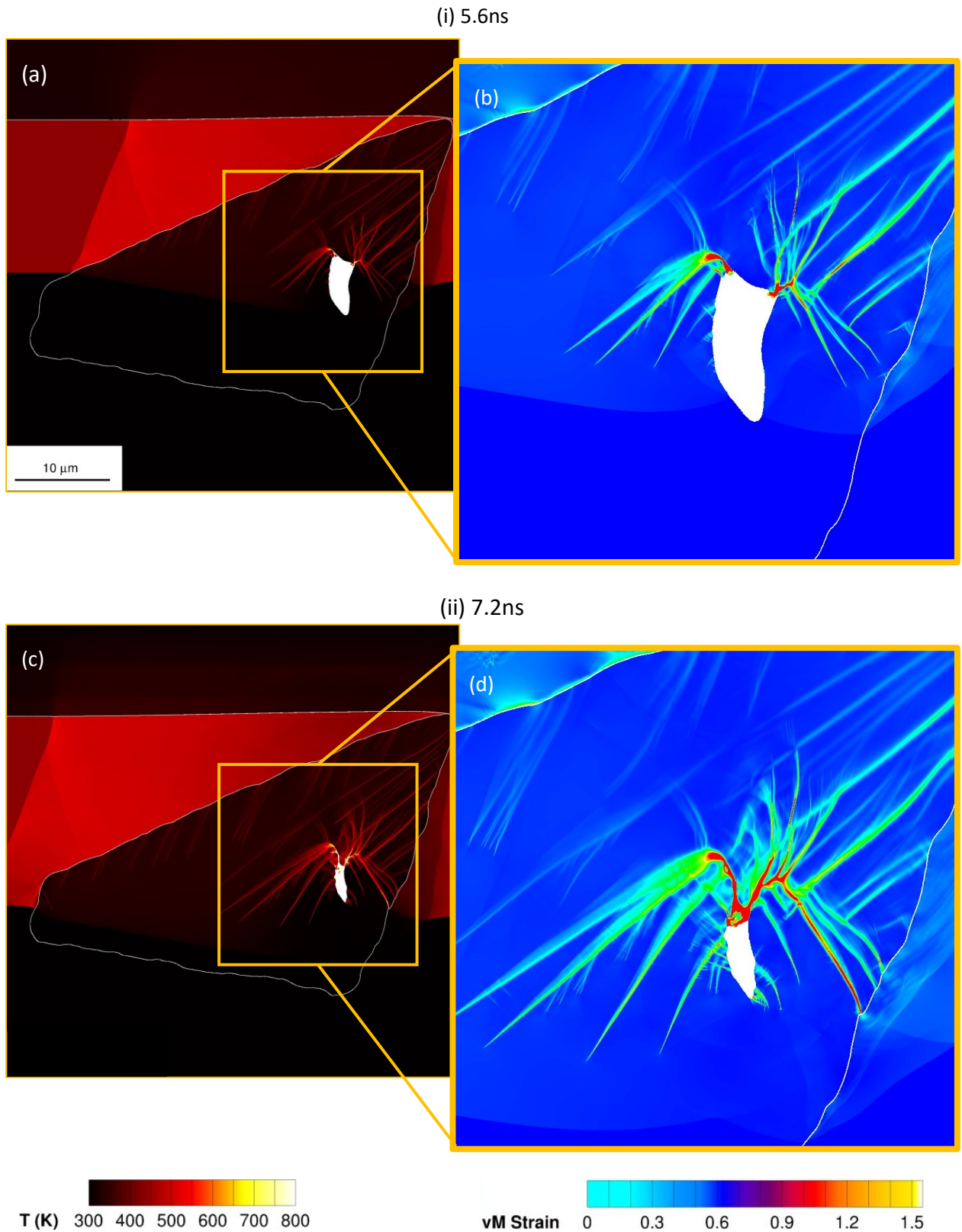

Figure 9. (a,c) Contour plots of temperature at 5.6ns and 7.2ns. (b,d) Zoomed-in views showing the von-Misses strain at the corresponding times. The impact from the Al flyer is explicitly tracked. Rate-dependent yield-stress model (M-JC) and temperature-dependent $C_v$ is used for HMX. 5$^{th}$-order WENO with Riemann-based GFM has been used.

This high-resolution study not only demonstrates the performance of the new models and numerical techniques but also opens the discussion for continuum calculations to consider real PBX microstructures beyond the use of ideal geometries such as circles/ellipses which do not show such rich heterogenous thermo-mechano-chemical phenomena. Further, the resolution for a continuum calculation employed here in combination with the atomistics-consistent material models enables the simulation of real crystal sizes and profiles, including internal defects and interfaces with binders, at a resolution approaching atomistic scales. In previous work [58, 133] we have shown that at such high resolutions and with the atomistics-consistent models, the mechanics of pore collapse and the temperature, shear stress and pressure fields generated are in close agreement with MD calculations. Since such MD calculations are likely to remain infeasible in the near term, at the dimensions of crystal shown in Figs. 8 & 9, the current high-resolution simulations are the best we can currently do to simulate what atomistic calculations may reveal if they were able to simulate such large systems.

## 4. Conclusions

We have presented an analysis of the effects of numerical treatments and material models on flyer-impact induced shock initiation of a model HMX-based PBX (single HMX crystal embedded in Estane binder). With the implementation of high-fidelity modeling aspects – boundary and interfacial treatment of material-material contact, order of accuracy of flux computations, and material strength model – we identify and assess the requirements for a multi-material continuum mechanics framework to best represent physical experiments.

The meso-scale calculations resulting from this study feature the following: 1) Crystal profiles are imported from nano-CT scans to have a faithful representation of geometries and heterogeneities such as surface undulations, internal pores and cracks. 2) The movement and impact by an Al flyer are simulated within the computational domain, conforming to experiments, instead of modeling the impact through a boundary condition at the domain boundary, 3) A high-order accurate $5^{th}$ order WENO scheme is used to integrate the fluxes and a Riemann-based GFM is employed in the sharp interface-tracking framework to accurately capture the material-material interactions. 4) An improved strength model for HMX: an isotropic modified Johnson-Cook type model that is calibrated against atomistics and shown to predict shear bands and hotspot temperatures with remarkable agreement to MD, 5) high mesh resolution has been employed to ensure that interfaces, shockwaves, and reaction fronts are adequately resolved.

Simulating the multi-material impact in lieu of approximating the impact through a boundary condition is a more direct approach that yields a more accurate thermo-mechanical response of the shocked PBX. Wave transmissions and reflections are naturally accounted for, which prevents the emergence of spurious waves and captures a physically consistent post-shock compression state. Although implementing a domain boundary condition may be computationally more tractable giving acceptable results in some cases, the strong shock regimes explored in this study revealed an unphysical pressure discrepancy several moments after the shock passage, which can skew hotspot growth behavior.

A similar influence is noticed in the choice of the specific heat model. A constant $C_v$ value is typically used in tandem with material models calibrated with atomistics (MD) which use the classical formulation [58]. The differences between the constant and temperature-dependent formulations are only minimal at high temperature regimes. However, at lower temperatures such as during post-shock relaxation, the constant $C_v$ model yields hotspots which are much cooler compared to the temperature-dependent $C_v$ model.

The elastic-perfectly-plastic (constant yield-stress) material model fails to capture shear effects. Importantly, simulations using this model captures a different mode of collapse of pores when compared with the rate-dependent model at low shock strength. The latter has been shown to better represent shear dominated pore collapse and shear bands at low velocity impacts, as well as hotspot temperatures at high shock strengths. In fact, the plasticity model is seen to have the strongest influence in conforming to experimentally documented hotspot temperatures for a strong shock initiation of a model HMX-based PBX.

The modelling aspects touched upon in this study serve to inform computational efforts which would involve large systems of crystal, binder, and defect/inclusion interactions at high speeds. Ongoing efforts that (require and) utilize the high-fidelity elements discussed in this work are focusing on establishing microstructural damage-sensitivity linkages [134] using direct numerical simulations, specifically looking towards capturing shock-to-detonation transition in PBXs, pressed HEs and composite energetic materials.

## Acknowledgements

The authors gratefully acknowledge support from AFOSR-MURI grant No. FA9550-19-1-0318, program manager: Capt. Derek Barbee. The authors extend their thanks to Prof. Dana D. Dlott and his lab group (UIUC, IL) for communicating nano-CT scans of the PBXs. Simulations were conducted on DoD HPC unclassified clusters – Onyx, Narwhal, and Carpenter.

## Data Availability Statement

The authors declare that the data supporting the findings of this study are available within the paper and can also be made available from the corresponding author upon reasonable request.

## Competing Interests

The authors have no competing interests to declare that are relevant to the content of this article.

**Appendix**

SI 1. Governing equations and constitutive relations

The continuum-scale conservation laws are as follows. The mass, momentum, and energy equations are cast in Eulerian form:

$$\frac{\partial \rho}{\partial t} + \frac{\partial (\rho u_i)}{\partial x_i} = 0 \tag{S1.1}$$

$$\frac{\partial (\rho u_i)}{\partial t} + \frac{\partial (\rho u_i u_j + p\delta_{ij})}{\partial x_j} = \frac{\partial S_{ij}}{\partial x_j} \tag{S1.2}$$

$$\frac{\partial \rho \left(e + \frac{1}{2} u_k u_k\right)}{\partial t} + \frac{\partial \left(\rho \left(e + \frac{1}{2} u_k u_k + p\right) u_j\right)}{\partial x_j} = \frac{\partial (u_i S_{ij})}{\partial x_j} + \dot{Q}_R + \chi \nabla^2 T. \tag{S1.3}$$

Here, $\chi$ is the thermal conductivity of the reaction mixture; $\rho$ is the material density; $u_i$ are the velocity components; $e$ is the specific internal energy; $T$ is the temperature; $p$ and $S_{ij}$ are respectively the volumetric and the deviatoric components of the Cauchy stress tensor: $\sigma_{ij} = -p\delta_{ij} + S_{ij}$, where $\delta_{ij}$ is the Kronecker delta, $\dot{Q}_R$ is the total heat release by chemical reaction.

The deviatoric stress tensor $S_{ij}$ in Eqs. (S1.2-1.3) is obtained from the following hypo-elastic constitutive model [1] in a rate (Prandtl-Reuss[2]) formulation:

$$\frac{\partial(\rho S_{ij})}{\partial t} + \frac{\partial(\rho S_{ij} u_k)}{\partial x_k} + \rho S_{ik}\Omega_{kj} - \rho\Omega_{ik} S_{kj} = 2\rho G\left(D_{ij}^d - D_{ij}^{pl}\right), \quad (S1.4)$$

where the left hand side is the objective Jaumann rate, $D_{ij}^d = \frac{1}{2}(u_{i,j} + u_{j,i}) - u_{k,k}\delta_{ij}/3$ is the deviatoric component of the strain-rate tensor, $\Omega_{ij} = \frac{1}{2}(u_{i,j} - u_{j,i})$ is the spin tensor, and $D_{ij}^{pl}$ is the plastic component of the deviatoric strain-rate tensor. Here, $G$ is the shear modulus of the material [3] which is represented by the MD-derived form (given in the following section), and $D_{ij}^{pl}$ is modeled assuming isotropic $J_2$ plasticity along with a von Mises yield criterion and an associated flow-rule. The consistency condition is enforced implicitly and $D_{ij}^{pl}$ is computed using a radial return algorithm[4], the details for which are provided in previous work[5].

Equation (S1.4) is solved using a two-step operator-splitting algorithm[6]. First, the deviatoric stress is evolved assuming a purely elastic deformation [$i.e.$, setting $D_{ij}^{pl} = 0$ in Eq. (S1.4)] in a predictor step. This is followed by a correction step to remap the predicted stress onto the yield surface using the radial return algorithm [4, 7, 8].

In the first step of solving for internal energy $e$, the specific internal energy $e^*$ is obtained from solving Eq. (S1.3) omitting $\dot{Q}_R$, followed by the temperature $T^*$ computed from the caloric equation of state:

$$T^*(\rho, e) = T_0 + (e^* - e_c)/C_V, \quad (S1.5)$$

where $T_0$ is the reference temperature (298 K) and $C_V$ is the isochoric specific heat.

To obtain the temperature from Eq. (S1.5) the "cold" (athermal) part of the specific internal energy $e_c$ is obtained from solving the following equation:

$$\rho\dot{e}_c = \rho\dot{e}_{c,hydro} + \rho\dot{e}_{el} + \rho\dot{e}_{c,pl}. \quad (S1.6)$$

where the subscripts $hydro$, $el$, and $pl$ denote respectively hydrodynamic, elastic, and plastic contributions. The hydrodynamic contribution to the athermal part of the internal energy $e_c$ is taken as:

$$\dot{e}_{c,hydro} = \frac{p_c(\rho)}{\rho^2}\dot{\rho} \quad (S1.7)$$

in which $p_c(\rho)$ is the cold compression contribution to pressure. $e_{c,hydro}$ is obtained from Eq. (S1.7) by integration,

$$e_{c,hydro} = e_0 + \int_{\rho_0}^{\rho} \frac{p_c(\rho)}{\rho^2} d\rho, \quad (S1.8)$$

where $e_0$ is the reference internal energy at $T_0$.

The plastic energy $e_{c,pl}$ contribution to cold specific internal energy is taken as a fraction of the plastic work via the Taylor-Quinney parameter $\beta$ as:

$$\frac{\partial \rho e_{c,pl}}{\partial t} + \frac{\partial (\rho e_{c,pl} u_j)}{\partial x_j} = (1-\beta) S_{ij} D_{ij}^{pl} \tag{S1.9a}$$

$$= (1-\beta) S_{vm} \dot{\varepsilon}_{pl}. \tag{S1.9b}$$

In the above, $S_{vm}$ is the effective (von Mises) stress and $\dot{\varepsilon}_{pl}$ is the effective plastic strain rate. In the last term on the R.H.S. of Eq. (S1.9a), the rate of plastic work $S_{ij} D_{ij}^{pl}$ is written as $S_{vm} \dot{\varepsilon}_{pl}$ for coaxial plasticity (J2 plasticity with Drucker's postulate[9]).

The elastic energy contribution $e_{c,el}$ is taken as the difference between the total deviatoric stress work and plastic work, and given by the following field equation in conservation form:

$$\frac{\partial \rho e_{c,el}}{\partial t} + \frac{\partial (\rho e_{c,el} u_j)}{\partial x_j} = S_{ij} D_{ij}^{el} \tag{S1.10a}$$

$$= S_{ij} D_{ij}^{d} - S_{vm} \dot{\varepsilon}_{pl}. \tag{S1.10b}$$

The discretization of the above equation for the elastic energy is performed in an identical manner as for the total internal energy $e$; $e_{c,el}$ is calculated at each time instant and grid point in the same manner as other field variables.

In the current work the Taylor-Quinney parameter is chosen to be unity due to lack of knowledge of the parameter for energetic crystals. The contribution of the plastic work to the temperature is found to be small for the cases studied in this paper.

In the second step of solving for total internal energy $e$, first, the temperature is updated using the total heat release by chemical reaction $\dot{Q}_R$.

$$\rho C_V \frac{T^{n+1} - T^*}{\delta t} = \dot{Q}_R \tag{S1.11}$$

In which $\dot{Q}_R = \sum_j^{N_{species}} h_j^s \dot{Y}_j$; $h_j^s$ is the sensible enthalpy of formation of $j^{th}$ species. The rise in temperature from chemical decomposition is used to update internal energy $e$ as:

$$e = e^* + C_V (T^{n+1} - T^*) \tag{S1.12}$$

as well as update pressure $p$ from the equation of state.

SI 2. Material models

*2.1 HMX*

For the inelastic response of HMX[10], the Johnson-Cook flow rule is employed, with the yield surface defined by

$$f = S_{vM} - \sigma_y = 0, \tag{S2.1}$$

where $S_{vM} \left(= \sqrt{\frac{3}{2}(S_{ij}S_{ij})}\right)$ is the effective (von Mises) stress and $\sigma_y$ is the yield stress of the material given by the Johnson-Cook form[11]:

$$\sigma_y = [A + B\varepsilon_{pl}^n][1 + C \ln(1 + \dot{\varepsilon}^*)]\left[1 - \left(\frac{T - T_{ref}}{T_m - T_{ref}}\right)^M\right]. \tag{S2.2}$$

Here, $\varepsilon_{pl}$ is the equivalent plastic strain, $\dot{\varepsilon}$ is the plastic strain rate, $\dot{\varepsilon}_0$ is a reference plastic strain rate, and $\dot{\varepsilon}^* = \dot{\varepsilon}_{pl}/\dot{\varepsilon}_0$; $T$ is the temperature and $T_{ref}$ is the reference temperature of the material (300 K); and $A$, $B$, $C$, $n$, and $M$ are Johnson-Cook model coefficients and exponents found in Das et al. (2021) [12].

The melt curve due to Kroonblawd and Austin [13], was used to denote the transition point for phase change to an inviscid liquid (modeled as $S_{ij}$=0):

$$T_m = T_{m0}\left(1 + \frac{p - p_{ref}}{a'}\right)^{1/c'}, \tag{S2.3}$$

where the reference pressure is $p_{ref} = 0$ and the quantities $a'$ and $c'$ are fitting parameters.

A pressure-dependent model for the shear modulus, with coefficients calibrated with MD calculations [14]:

$$G(p) = -2.0543 \times 10^{-11} p^2 + 1.7531 p + 7.89 \times 10^9 \text{ Pa} \tag{S2.4}$$

In Eq. (3) the pressure is expressed in Pascal units.

If the material is stressed beyond a threshold, marked by the following conditions being fulfilled: (1) the von Mises stress becomes larger than a threshold value, $S_{vM} > \sigma_e^c$, where $S_{vM} = \left\{\frac{3}{2}(S_{ij}S_{ij})\right\}^{\frac{1}{2}}$, and (2) the equivalent plastic strain increases above a limit, $\varepsilon_{pl} > \varepsilon_{eq}^c$, then the flow-stress model is replaced with:

$$\sigma_y^* = \frac{\sigma_y + \sigma_{flow}^\infty}{2} + \frac{\sigma_{flow}^\infty - \sigma_y}{2}\tanh(\varepsilon_{pl}). \tag{S2.5}$$

In Eq. S27, $\sigma_y$ is computed with Eq. S24, and the stress threshold is a pressure-dependent critical value defined by:

$$\sigma_e^c = -0.003425p^2 + 0.1557p + 0.355 \text{ GPa}. \tag{S2.6}$$

In Eq. (9) the pressure is expressed in Gigapascal units. $\sigma_{flow}^\infty$ in Eq. (8) is written as:

$$\sigma_{flow}^{\infty} = \sqrt{\frac{3}{2}} A_{flow}(p)(\sqrt{3}\dot{\varepsilon}_{pl})^{0.1} |1 - B_{flow}(p)T| f(p,T). \tag{S2.7}$$

This form accounts for the power-law dependence of the resolved shear stress on the resolved shear-strain rate and for the linear dependence of the resolved flow stress on temperature. The functions $A_{flow}(p)$ and $B_{flow}(p)$ are calibrated based on atomistic simulation results and can be written as[15],

$$A_{flow}(p) = 0.023 \text{ GPa} + \frac{p}{100}, \tag{S2.8}$$

$$B_{flow}(p) = 0.001 \left[1 - 0.75\left(1 - e^{-\frac{p}{6\,GPa}}\right)\right], \tag{S2.9}$$

resulting in units of GPa for $\sigma_{flow}^{\infty}$. The 'restriction factor' $f(p,T)$ in Eq. (10) is included to ensure that $\sigma_{flow}^{\infty}$ is non-negative at all pressures and temperatures. Specifically, $\sigma_{flow}^{\infty}$ is restricted to be positive by defining $f(p,T)$ as

$$f(p,T) = \frac{1}{2}\left[1 - \tanh\left(\frac{T + 100 - \frac{1}{B(p)}}{50}\right)\right], \tag{S2.10}$$

where $T$ is expressed in kelvins.

The threshold value $\varepsilon_{eq}^c$ was set in the present work to be just above the value of the post-shock strain far away from the pore, $\varepsilon_{pl}^{\infty}$; that is, $\varepsilon_{eq}^c = \varepsilon_{pl}^{\infty} + \delta\varepsilon$, in which $\delta\epsilon$ is a small increment of plastic strain.

The Mie-Grüneisen EOS is employed for HMX in conjunction with the 3rd-order Birch-Murnaghan form for the cold compression contribution to pressure,

$$p_c = \frac{3}{2}K_0\left[\left(\frac{\rho_0}{\rho}\right)^{\frac{-7}{3}} - \left(\frac{\rho_0}{\rho}\right)^{\frac{-5}{3}}\right]\left[1 + \frac{3}{4}(K_0' - 4)\left\{\left(\frac{\rho_0}{\rho}\right)^{\frac{-2}{3}} - 1\right\}\right], \tag{S2.11}$$

where the coefficients $K_0$ and $K_0'$ are the isothermal bulk modulus and its initial pressure derivative, respectively, and $V_0$ is the reference volume at (300 K, 1 atm).

The specific heat $C_V$ in Eq. (S1.5) is a key thermophysical quantity for hotspot formation, as the temperature reached after pore collapse is influenced by its value. In MD calculations, due to the inherent classical approximation, the effective $C_V$ in the system is practically constant. On the other hand, in continuum calculations a temperature-dependent $C_V$ [16] is commonly used. Both the formulations of $C_V$ has been tested in this work. For simulations that maintain strict consistency between the continuum and MD models, the constant of $2359 Jkg^{-1}K^{-1}$ is used corresponding to the value implicit in the MD simulations[3]. The temperature-dependent $C_V$ is of the form:

$$C_V = \frac{\tilde{T}^3}{c_0 + c_1\tilde{T} + c_2\tilde{T}^2 + c_3\tilde{T}^3}, \tag{S2.12}$$

Where $\tilde{T} = T/\theta(V)$, where $V = 1/\rho$. Here, $\theta(V)$ is the Debye temperature and has the form

$$\theta(V) = \theta_0 \left(\frac{V_0}{V}\right)^a \exp\left[\frac{b(V_0 - V)}{V}\right]. \tag{S2.13}$$

The constants $V_0, K_0, K_0', a, b, c_{0-3}, \theta_0$ can be found in Das et al. [12].

## 2.2 Estane

*Estane is modeled as an inert viscoelastic material. Stress is governed by the equation for the evolution of deviatoric stresses given in Eq (S1.4). Beyond a constant melt temperature of 473K, the material is approximated as an inviscid liquid. The shear response for Estane has been modeled by a constant shear modulus $G = 1$GPa.*

The temperature in the binder is computed from the specific internal energy:

$$T = T_0 + \frac{(e - e_0)}{C_V} \tag{S2.14}$$

where the reference temperature in binder, $T_0$ is 298.14 K and the reference specific internal energy of binder, and $e_0 = 0$ J/kg. Other material and EOS parameters for Estane are given in Table S1.

## 2.3 Aluminum

*Aluminum is modeled as an inert elastoplastic material, with a constant melt temperature, shear modulus, specific heat coefficient and thermal conductivity.* Other material and EOS parameters for Aluminum are given in Table S1.

Table S1: EOS and material parameters of materials used in this work [17].

| Material | Reference density $\rho_0 \left(\frac{kg}{m^3}\right)$ | Isochoric specific heat coefficient $C_V \left(\frac{J}{Kg.K}\right)$ | Thermal conductivity $\chi \left(\frac{W}{m.K}\right)$ | Gamma $\Gamma_0$ | Bulk sound speed $C_0 \left(\frac{m}{s}\right)$ | Slope of $U_s$-$U_p$ (for MG EOS) $s$ |
|---|---|---|---|---|---|---|
| HMX | 1900 | Max($f$(T), 967.571) | See Table S2 | 1.1 | 2650.0 | - |
| Estane | 1190 | 2076 | 0.148 | 1 | 1750 | 2.2 |
| Aluminum | 2785 | 896 | 166.9 | 1.99 | 5386 | 1.339 |

Table S2. Specific thermal conductivity for all four species from the Tarver et al. three-equation model [1] for various temperatures.

| Thermal conductivity $\chi$ (cal/cm s K) | HMX | Fragments | Intermediate gases | Final gases |
|---|---|---|---|---|
| 293 K | $1.23 \times 10^{-3}$ | $6.5 \times 10^{-4}$ | $1 \times 10^{-4}$ | $1 \times 10^{-4}$ |
| 433 K | $9.7 \times 10^{-4}$ | $5 \times 10^{-4}$ | $1 \times 10^{-4}$ | $1 \times 10^{-4}$ |
| 533 K | $8.1 \times 10^{-4}$ | $4 \times 10^{-4}$ | $1 \times 10^{-4}$ | $1 \times 10^{-4}$ |
| 623 K | $7 \times 10^{-4}$ | $3 \times 10^{-4}$ | $1 \times 10^{-4}$ | $1 \times 10^{-4}$ |

SI 3. Sharp-interface tracking using levelsets

Levelsets are used to represent the embedded material objects[18]. The zero-level of the signed distance function field $\phi$ sharply delineates materials interfaces. The levelset field is defined only in a narrow band and evolved using the level set advection equation[19]:

$$\frac{\partial \phi}{\partial t} + u_{norm} \cdot \nabla \phi = 0 \tag{S3.1}$$

Eq. (23) is solved using the 5th -order WENO scheme for spatial discretization and 3rd-order Runge-Kutta for time integration. The levelset velocity $u_{norm}$ is defined only on the object boundary by the physics of the material interface and extended into the narrow band by solving the Eikonal equation[20]:

$$\frac{\partial \psi}{\partial t} + u_{ext} \cdot \nabla \psi = 0 \tag{S3.2}$$

$\psi$ is a variable to be extrapolated such as the interface velocity components. The extension velocity is given by[18]:

$$u_{ext} = sign(\phi) \frac{\nabla \phi}{|\nabla \phi|} \tag{S3.3}$$

The advection of the level set field does not preserve the properties of the signed distance function. Standard reinitialization is performed at regular intervals to return the levelset field to a signed distance function such that $|\vec{\nabla} \phi| = 1$. Reinitialization is done by solving the following equation to steady state:

$$\frac{\partial \phi}{\partial \tau} + w(|\nabla \phi| - 1) = 0 \tag{S3.4}$$

Where $\tau$ is a pseudo-time and $w = sign(\phi_o)$ where $\phi_o$ is the level set field at the start of the reinitialization procedure. Further details of the implementation of the sharp interface tracking method in SCIMITAR3D can be found in [5].

# SI 4. Effect of pulse width on hotspot growth for a shock-pulse boundary condition – based modeling

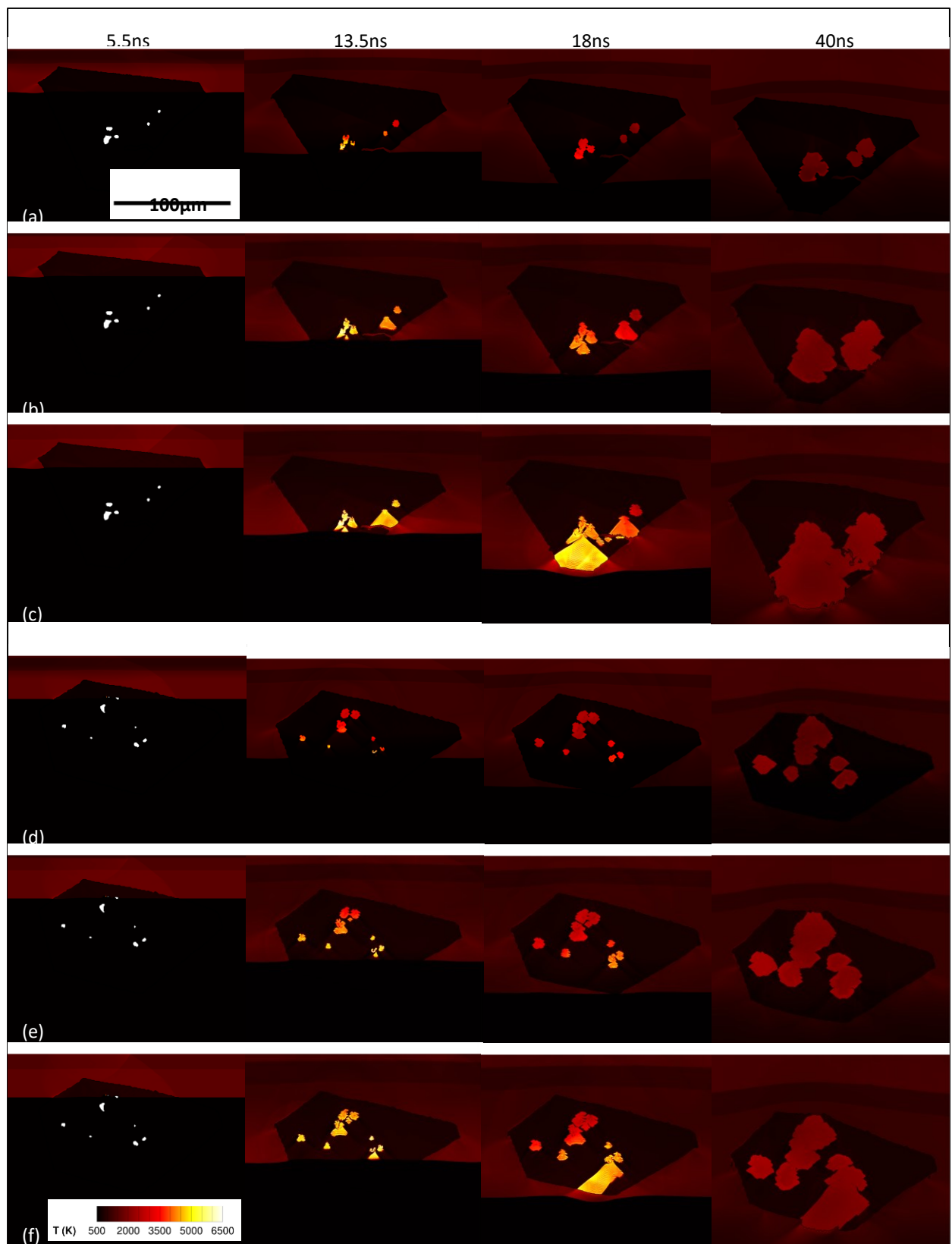

Figure S1. Temperature contour plots from simulation of PBX sample, consisting of an HMX crystal embedded in binder, subjected to a shock-pulse with Up=2.3km/s at the binder surface. Crystal '1' with pulse duration (a) 4ns, (b) 5.3ns, (c) 7ns. Crystal '2' with pulse duration (d) 4ns. (e) 5.3ns. (f) 7ns.